\documentclass[showpacs,onecolumn,preprintnumbers,amsmath,amsfonts,amssymb,floatfix,aps,superscriptaddress]{revtex4}
\usepackage{amsmath}
\usepackage{amsfonts}
\usepackage{amssymb}
\usepackage{graphicx}
\usepackage{epsfig}
\usepackage{siunitx}
\usepackage[hang,nooneline]{subfigure}
\usepackage{hyperref}

\newcounter{fig}
\newcommand{\lbfig}[1]{\refstepcounter{fig}           .
}

\begin{document}

\title{Extreme events in near integrable lattices}

\author{C. Hoffmann}
\affiliation{Department of Mathematics and Statistics, University of Massachusetts
Amherst, Amherst, MA 01003-4515, USA}
\affiliation{Institut f\"ur Physik, Universit\"at Oldenburg,
D-26111 Oldenburg, Germany}

\author{E. G. Charalampidis}
\affiliation{Department of Mathematics and Statistics, University of Massachusetts
Amherst, Amherst, MA 01003-4515, USA}

\author{D. J. Frantzeskakis}
\affiliation{Department of Physics, National and Kapodistrian University of Athens, Panepistimiopolis, 
Zografos, Athens 15784, Greece}

\author{P. G. Kevrekidis}
\affiliation{Department of Mathematics and Statistics, University of Massachusetts
Amherst, Amherst, MA 01003-4515, USA}

\begin{abstract}
  In the present work, we 
  examine the potential robustness of 
  extreme wave events associated with large
  amplitude fluctuations of the Peregrine soliton type, upon departure 
  from the integrable 
  analogue of the
  discrete nonlinear Schr{\"o}dinger (DNLS) equation, namely
  the Ablowitz-Ladik (AL) model. Our model of choice will be the
  so-called Salerno model, which interpolates between the
  AL and the DNLS models. We find that rogue wave 
  events essentially are drastically distorted even for
  very slight perturbations of the homotopic parameter connecting
  the two models off of the integrable limit. Our results
  suggest that the Peregrine soliton structure is a rather sensitive
  feature of the integrable limit, which may not persist under
  ``generic'' perturbations of the limiting integrable case.
  \end{abstract}

\date{\today}

\maketitle 

\begin{section}{Introduction}

  The study of phenomena associated with extreme events
  and rogue waves has gained substantial traction over
  the last few years~\cite{k2a,k2b,k2c,k2d}. This can
  largely be attributed to the development of experimental
  settings in a variety of fields where the relevant coherent
  structures can be systematically created and observed.
  These fields range from superfluid helium~\cite{He}
  to hydrodynamics~\cite{hydro,hydro2,hydro3}, and
  from nonlinear optics~\cite{opt1,opt2,opt3,opt4,opt5,laser} and 
  plasmas~\cite{plasma} to Faraday surface ripples~\cite{fsr}
  and parametrically driven capillary waves \cite{cap}.
  Experimental efforts have, in part, been motivated by --and also inspired-- 
  numerous theoretical investigations, mainly concerning 
  variants of the nonlinear Schr{\"o}dinger (NLS) equation.
  The theoretical activity has now been summarized in
  numerous reviews~\cite{yan_rev,solli2,onorato}, as
  well as books~\cite{k2a,k2b,k2c,k2d}.

  One of the significant aspects of the investigation of
  extreme wave events has to do with the structural form
  that these events assume, and perhaps especially with their robustness 
  in NLS and related models.
  The seminal works of Peregrine~\cite{H_Peregrine}, Kuznetsov~\cite{kuz},
  Ma~\cite{ma}, and Akhmediev~\cite{akh}, as well as of Dysthe and Trulsen~\cite{dt}, 
  have provided a framework of study of relevant coherent structures, either periodic
  in space (such as the Akhmediev breather) or periodic in
  time (such as the Kuznetsov-Ma breather) 
  or, most notably, localized in space-time, 
  as the Peregrine soliton~\cite{H_Peregrine}.
  A question then emerges about whether these entities survive
  model perturbations and/or emerge under generic classes of
  initial conditions. Admittedly, the latter question has only
  been partially addressed. For instance, in a class of perturbations
  involving Hirota-model variants (such as third-order dispersion
  and self-steepening terms), a perturbed, still algebraically decaying
  variant of the Peregrine soliton was obtained (however its persistence
  was not ensured to all orders in the perturbation)~\cite{devine}.
  Moreover, adiabatic approximations~\cite{cagnon} and
  perturbed inverse scattering approaches~\cite{KalimerisGarnier}
  have considered the stability of Kuznetsov-Ma (KM) solitons indicating
  their potential robustness against dispersive but non-robustness
  against dissipative perturbations. A different perspective on
  the emergence of localized phenomena in space-time was given
  by the work of~\cite{calinibook}, where it was argued that
  the proximity of such solutions to chaotic states (in more
  elaborate, non-integrable models) appears to increase the
  occurrence of 
  extreme events. Other works have focused on
  the stability properties of solutions~\cite{vangorder,abdul1,abdul2};
  however, there is an ambiguity associated with the
  time-dependent nature of the solutions.
  A natural setup for performing stability studies is, arguably,
  the Floquet analysis of the
  time-periodic KM solution~\cite{usfloquet}.

  In a recent work~\cite{stathis}, 
  a different type of ``genericity'' 
  of these solutions was considered: the emergence of extreme events 
  stemming from simple --yet typical--  
  Gaussian initial data, under a phenomenon called gradient catastrophe that has
  been explained in the pioneering work of~\cite{bertola}. In
  particular, it was proposed that in the semiclassical limit of
  the NLS model, such initial data will lead to the formation
  of an array of essentially identical (up to small corrections)
  Peregrine soliton-like structures, which emerge at the poles of
  the so-called tritronqu{\'e}e solution of the Pain{\'e}v{\'e} I
  equation. The importance of these findings is underscored
  by the fact that very recently the universality of this
  emergence of the Peregrine soliton in the semiclassical
  focusing dynamics of the NLS model has been manifested
  experimentally~\cite{suret}. Furthermore, recently such Peregrine
  waveforms were also found to spontaneously emerge as a result
  of the interaction of dispersive shock waves in~\cite{khamis}.

  Our aim in the present work is to explore the relevance of
  Peregrine-soliton type solutions in spatially discrete systems, 
  i.e., in nonlinear dynamical lattices. 
  There has been an amount of work in this context as well. 
  In particular,
  it has been established that a Peregrine-like solution, strongly
  reminiscent of its continuum sibling exists~\cite{akhmdis} in the context of
  the completely integrable discrete version of the NLS equation, the so-called
  Ablowitz-Ladik (AL) model~\cite{AL1,AL2}.
  In fact, subsequent work has established the systematic construction
  of higher-order such solutions~\cite{ohtayang}.
  However, it is also well-known
  that while the AL model is useful for the consideration of numerous
  perturbative calculations involving single discrete solitons~\cite{cai},
  their stability~\cite{kapit} and 
  their collisional
  dynamics~\cite{dmitriev}, it is not of direct relevance to experimental
  settings. On the contrary, the quintessential discrete model of
  relevance, both to nonlinear optics (in the context of arrays of coupled waveguides) 
  and to atomic Bose-Einstein condensates (BECs) confined in optical lattices, 
  is the discrete nonlinear Schr{\"o}dinger (DNLS) equation~\cite{dnlsbook}.
  Hence, our considerations herein will involve departing in a
  systematic way from the AL model and approaching the DNLS one.
  This will be done through the 
  Salerno model~\cite{salerno}, 
  interpolating between the two limits.

  We consider a Gaussian initial profile (as a generic waveform)
  and examine a two-parametric variation. In particular, on the side of varying the
  initial condition parameters, we examine the effect of changing 
  the variance of the initial condition (IC). 
  Here, using a large variance
  places us within the so-called semi-classical regime~\cite{bertola},
  where we may expect analogously to the continuum case of~\cite{stathis}
  to observe Peregrine soliton like structures. On the other hand, at the level
  of varying the model, we consider changes of a homotopic parameter 
  as extending from the AL limit all the way to the DNLS
  one, and examine a wide variety of cases in between. Our main
  observation is that at the AL limit, we identify Peregrine 
  structures and even a space-time evolution featuring 
  the emergence of a ``Christmas-tree''-like pattern, 
  analogous to the continuum case~\cite{stathis}, 
  as an apparent discrete emulation
  of the gradient catastrophe phenomenology of~\cite{bertola}.
  Nevertheless, this appears to be --in some sense-- a singular case,
  in that as soon as we depart from this integrable limit, 
  the prevalent dynamical structures appear to consist of
  persistent or breathing in time discrete
  solitonic entities (discussed at length in the context
  of DNLS models~\cite{dnlsbook}), rather than of Peregrine-like
  patterns. It is intriguing to point out that a similar conclusion
  (a propensity towards freak waves near the integrable limit)
  had emerged through the important statistical analysis
  of Ref.~\cite{tsironis}.
  In fact, our observations
  lead us to conjecture that no direct analogue of the Peregrine
  soliton exists in the DNLS model, although one exists in its
  continuum limit, as well as in its integrable discrete sibling. A dynamical
  systems analysis that would tackle this persistence problem
  would be of paramount importance for future work.

  Our presentation is structured as follows. In section~II, we
  present the relevant mathematical model(s) and the corresponding
  prototypical solutions. In section~III, we establish the corresponding
  numerical results and comment on the relevant observations.
  Finally, in section~IV, we present a summary of our findings and
  provide some suggestions for future work.

\end{section}
  
\begin{section}{The model}
The model of interest originates from the focusing 
NLS equation, written in dimensionless form as follows:
\begin{equation}
i\partial_{t}u =-\frac{1}{2}\partial_{x}^2 u -|u|^2u,
\label{nlse_1d}
\end{equation}
where $u(x,t)\in \mathbb{C}$ is the wave function. Next,
discrete realizations of Eq.~\eqref{nlse_1d} can be obtained, e.g., 
by replacing the (continuous) dependent variable $u(x,t)$ 
with $u_{n}(t)\doteq u(x_{n},t)$ ($x_{n}=-L+nh$ with $L$
the grid's half-width) and the second partial derivative 
with its central finite difference operator. This way, we obtain the
discrete NLS (DNLS) equation: 
%
\begin{equation}
i\dot{u}_{n} =-\frac{1}{2h^{2}}\left(u_{n+1}-2u_{n}+u_{n-1}\right) -|u_{n}|^2u_{n},%
\quad n\in\mathbb{Z},
\label{dnls_1}
\end{equation}
where overdot stands for differentiation with respect to time and,  
hereafter, we will set the lattice spacing $h=1$. 
Details on the derivation and physical origin of the DNLS, e.g., 
in coupled optical waveguides and in BECs confined in optical lattices, as well 
its discrete soliton solutions, can be found in the review~\cite{dnlsbook}. 
%
%

A different discretization of Eq.~\eqref{nlse_1d}
can be obtained by discretizing the field value multiplying the
square modulus in Eq.~\eqref{dnls_1} as
$u_{n}\doteq(u_{n+1}+u_{n-1})/2$. The resulting discrete lattice model is the AL
model~\cite{AL1,AL2}, which is of the form:
\begin{equation}
i\dot{u}_{n} =-\frac{1}{2}\left(u_{n+1}-2u_{n}+u_{n-1}\right)-%
\frac{1}{2}|u_{n}|^2\left(u_{n+1}+u_{n-1}\right).
\label{al_1}
\end{equation}

To interpolate between the DNLS and the AL models, we
introduce a real parameter $\mu\in[0,1]$. Then, we can write the following  
``tunable'' discrete lattice system: 
%
\begin{equation}
i\dot{u}_{n} =-\frac{1}{2}\left(u_{n+1}-2u_{n}+u_{n-1}\right)-%
\mu|u_{n}|^{2}u_{n}-
\frac{1}{2}\left(1-\mu\right)|u_{n}|^2\left(u_{n+1}+u_{n-1}\right),
\label{gener_model}
\end{equation}
which corresponds to the DNLS and AL 
models for $\mu=1$ and $\mu=0$, respectively. 
This generalized Salerno model~\cite{salerno} of Eq.~\eqref{al_1} will be the
focal point of our subsequent
numerical investigations. 

Here, it should be noted 
that the AL model supports rational solutions of the rogue wave type: 
in particular, the first-order such rational solution 
of the AL system 
is of the form~\cite{akhmdis}: 
\begin{equation}
u_{n}=U_{n}e^{i\phi}, \quad U_{n}(t)=\left(\frac{4q\left(1+q^{2}\right)%
\left(1+2iq^{2}t\right)}{1+4q^{2}n^{2}+4q^{4}\left(1+q^{2}\right)t^{2}}-q\right)e^{iq^{2}t},
\label{dis_per}
\end{equation}
with $q$ and $\phi$ being a real background amplitude and (arbitrary) 
phase, respectively. It is to that solution that we will compare our 
findings, especially so in computations associated with the AL limit.
In that light, the background amplitude $q$ will be utilized as a
fitting parameter to obtain the ``best-fit'' Peregrine soliton.

Our goal is to study the initial value problem (IVP) which consists
of Eq.~\eqref{gener_model} (for various values of $\mu\in[0,1]$) and
Gaussian initial data, of the form: 
\begin{equation}
u_{n}(t=0)=e^{-n^{2}/2\sigma^{2}}, 
\label{dis_gauss}
\end{equation}
where $\sigma$ characterizes the Gaussian's width. In a vein reminiscent
of the work of~\cite{stathis}, we are interested in identifying parametric
regimes of both $\sigma$ and $\mu$, such that extreme events (or
fundamental solitons) at the discrete level can be obtained. 

\end{section}

\begin{section}{Numerical Results}

  Our exposition of numerical results commences with the
  case of
Eq.~\eqref{gener_model} for $\mu=0$, i.e., 
the AL model. Specifically, periodic
boundary conditions are employed on a lattice containing 
$N=512$ points. At first, and as a benchmark case, we 
initialize the dynamics with the Peregrine soliton given 
by Eq.~\eqref{dis_per} at $t=-30$, with $q=1/3$ and $\phi=0$; 
the results are presented in Fig.~\ref{fig1}. It can be discerned
from this figure that the formation of the Peregrine soliton
is clearly evident (see the left panel therein), as well as that 
the numerics compare well with the exact results (see the 
right panel of the same figure, where densities at $t=0$ of 
the exact and numerical solutions are presented).

\begin{figure}[tbp]
\begin{center}
\includegraphics[height=.18\textheight, angle =0]{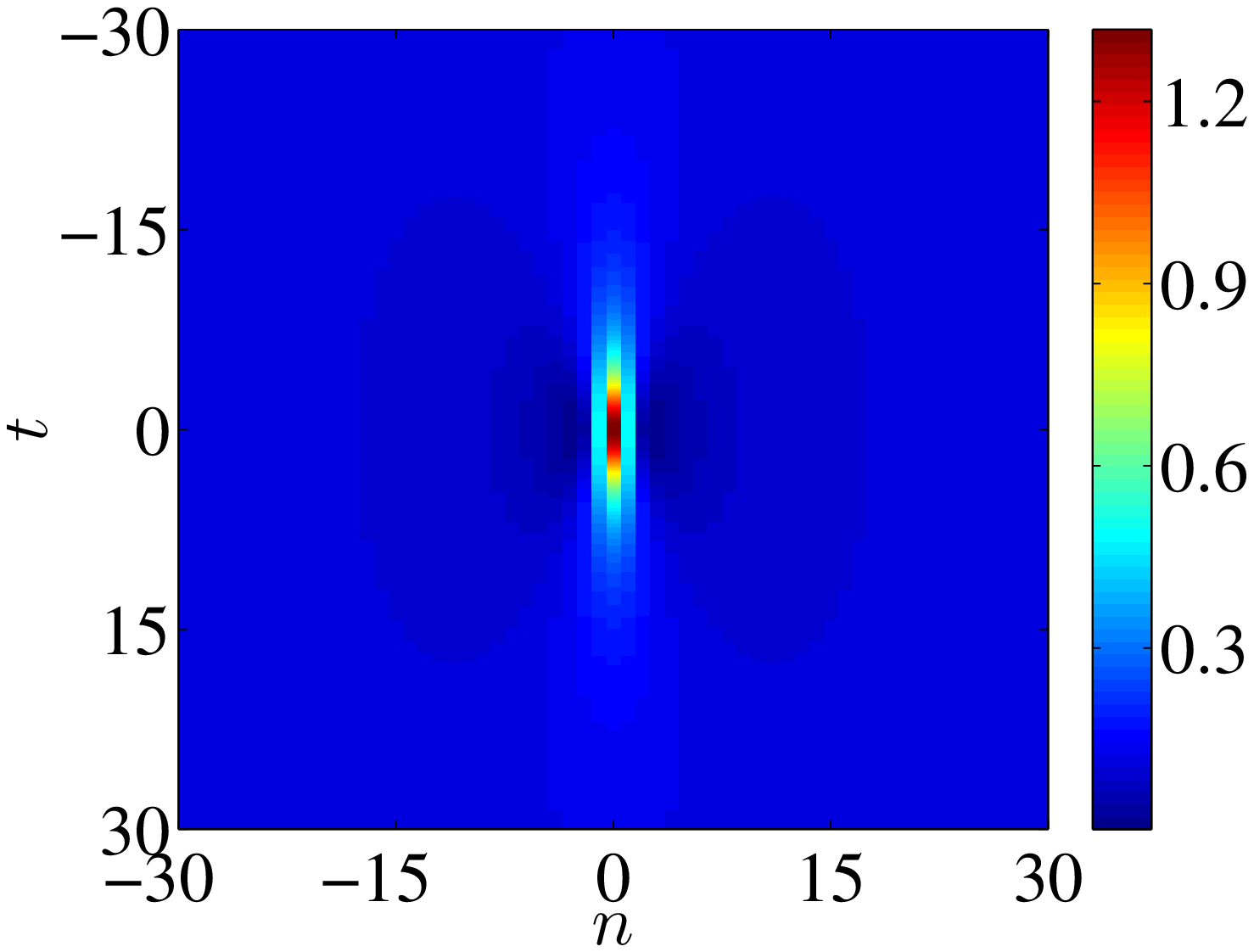}
\includegraphics[height=.18\textheight, angle =0]{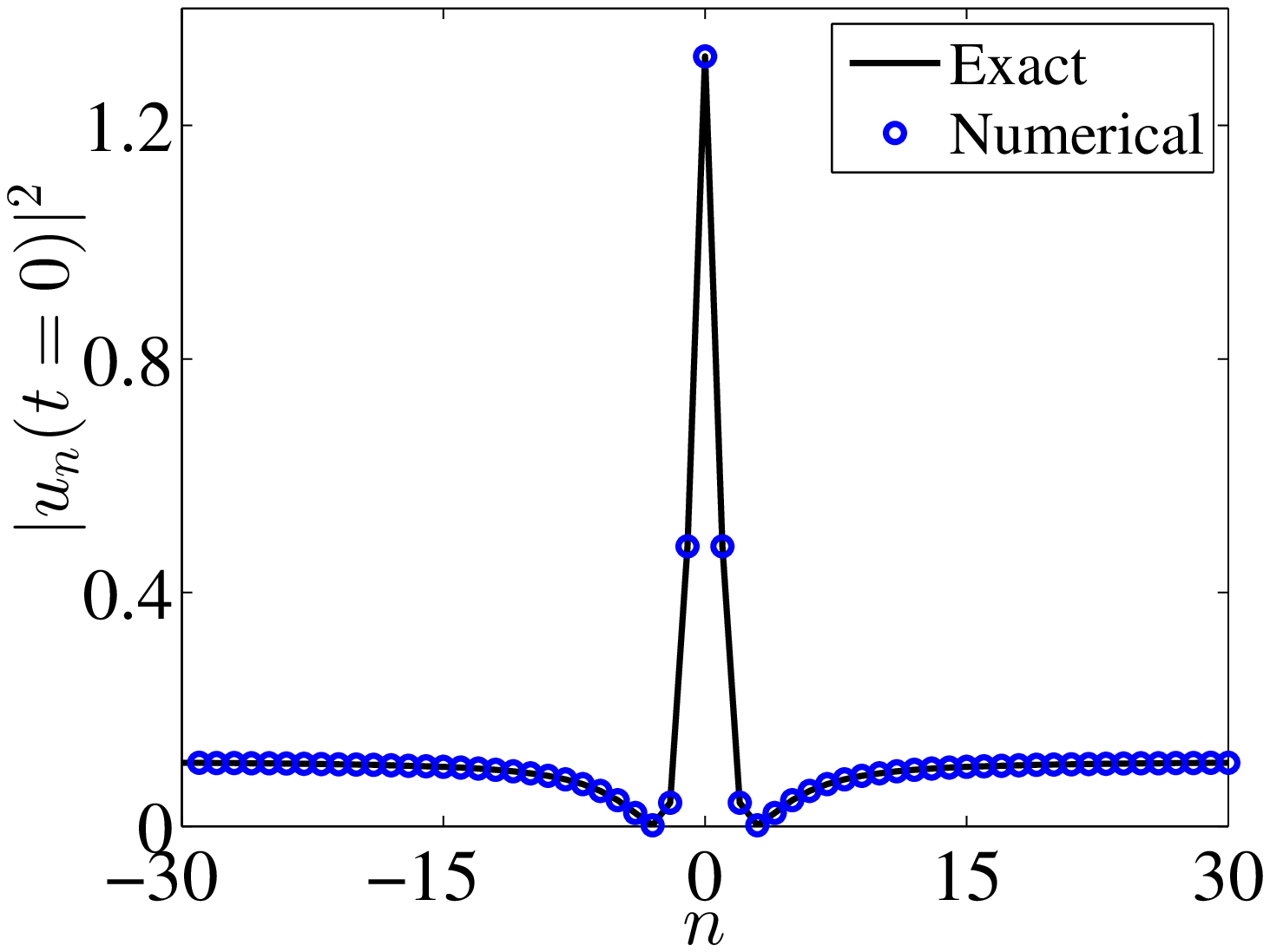}
\end{center}
\caption{
(Color online) Numerical results for the Ablowitz-Ladik (AL)
model with Peregrine soliton initial data, given by Eq.~\eqref{dis_per}. The
spatiotemporal evolution of the density $|u_{n}|^{2}$ is shown
in the left panel, whereas the spatial distribution of the density
evaluated at $t=0$ is presented in the right panel, with blue open circles; 
the exact solution is plotted too, with a black
solid line, for comparison. }
\label{fig1}
\end{figure}

Next, and still in the realm of the AL model, we initialize 
the dynamics using the discrete Gaussian wavepacket given 
by Eq.~\eqref{dis_gauss}. Representative results with $\sigma=30$
and $\sigma=15$ are presented in the top and bottom rows
of Fig.~\ref{fig2}, respectively. 
For large 
values of the Gaussian's width, extreme events can be seen to form 
(see also Ref.~\cite{stathis} for the continuum NLS case) 
in a way strongly reminiscent of the Peregrine solitons 
at the discrete level (see the right panels of Fig.~\ref{fig2}).
Some observations are in order here. On the one hand, it
appears evident that gradient catastrophe type phenomena
are still present in the discrete case, in analogy with the
continuum one~\cite{bertola}. Moreover, extreme intensities
are achieved due to the ability of the discrete problem
to collect a large fraction of the original mass at a single site
(in the continuum 1D problem, the mass conservation and absence
of collapse do not favor such an extreme accumulation).
The individual constituents of the resulting structure, much
reminiscent of the ``Christmas tree'' pattern of Ref.~\cite{stathis}
are evidently well approximated by a Peregrine soliton. As shown
in the right panel near the core, a Peregrine pattern of the
same maximal intensity allows to closely approximate the
core of the wave, although obviously the different asymptotics
of the Gaussian vs. the exact Peregrine necessitate a
deviation between the two far enough from the center.
Lastly, other things partially differ from the generic continuum
picture drawn through the insights of Ref.~\cite{bertola}. For instance,
the discrete nature of space (not allowing the solitons to be
centered exactly at the poles of the tritronqu{\'e}e solutions)
leads to a substantial variation of amplitude between the
different peaks.

\begin{figure}[tbp]
\begin{center}
\includegraphics[height=.18\textheight, angle =0]{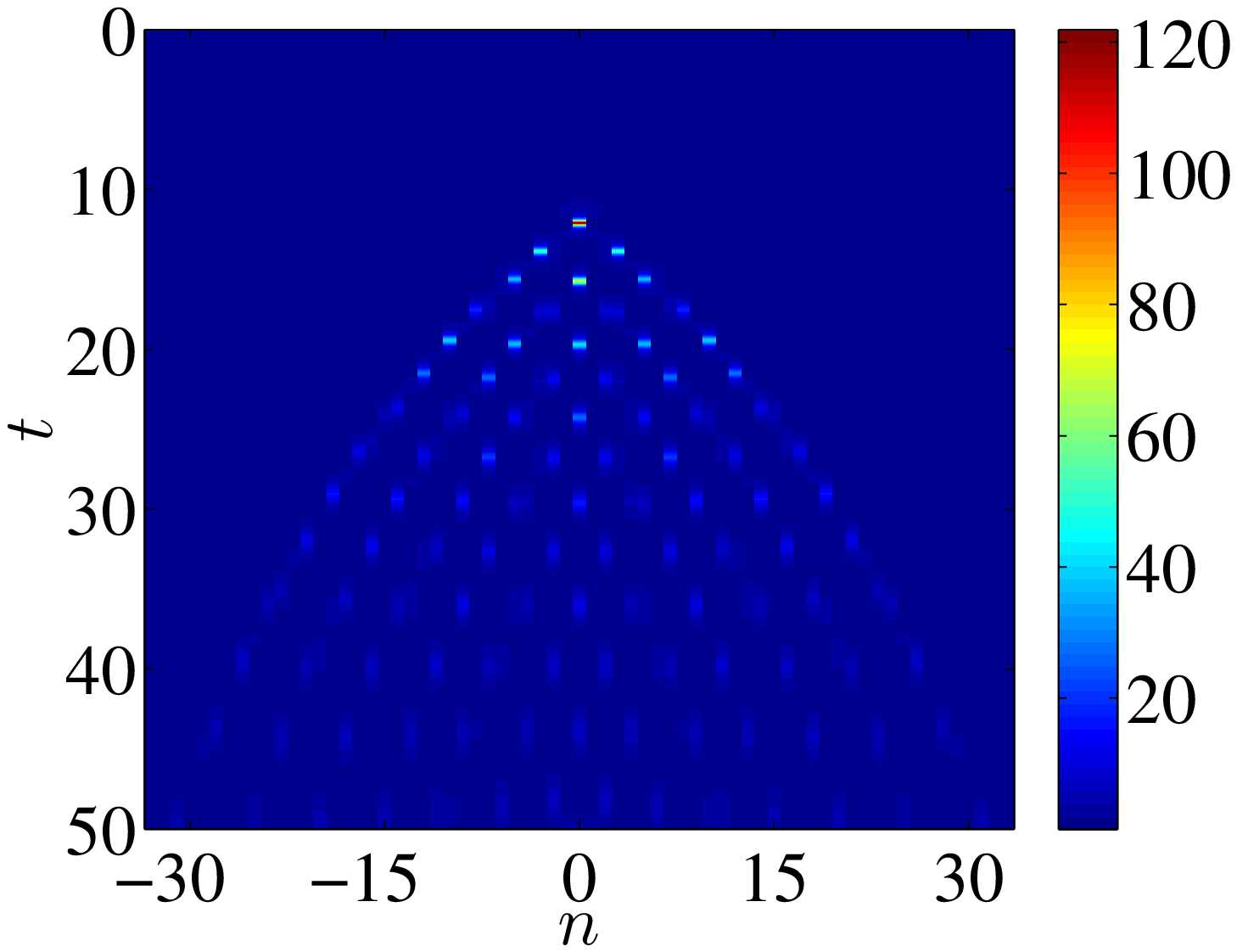}
\includegraphics[height=.18\textheight, angle =0]{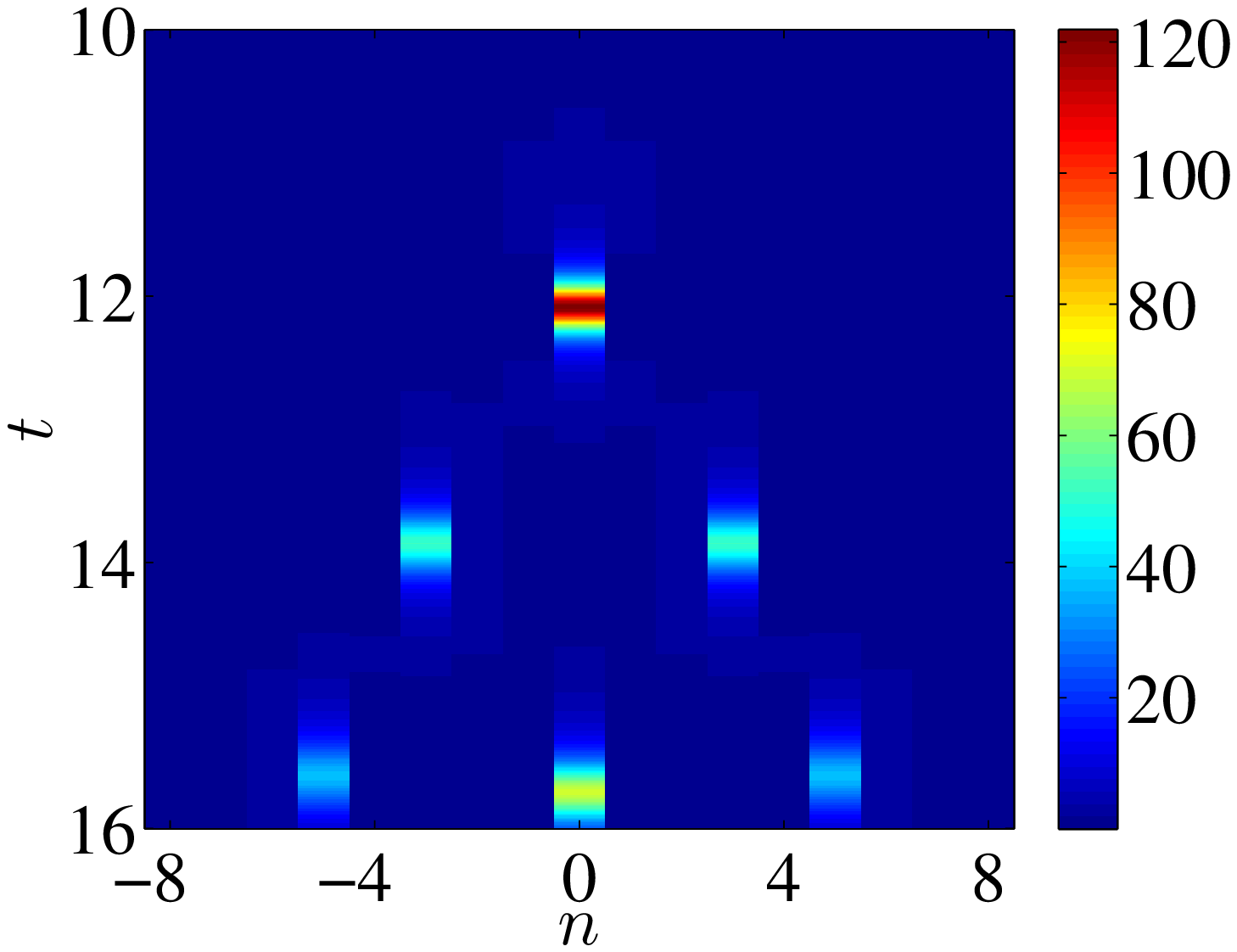}
\includegraphics[height=.18\textheight, angle =0]{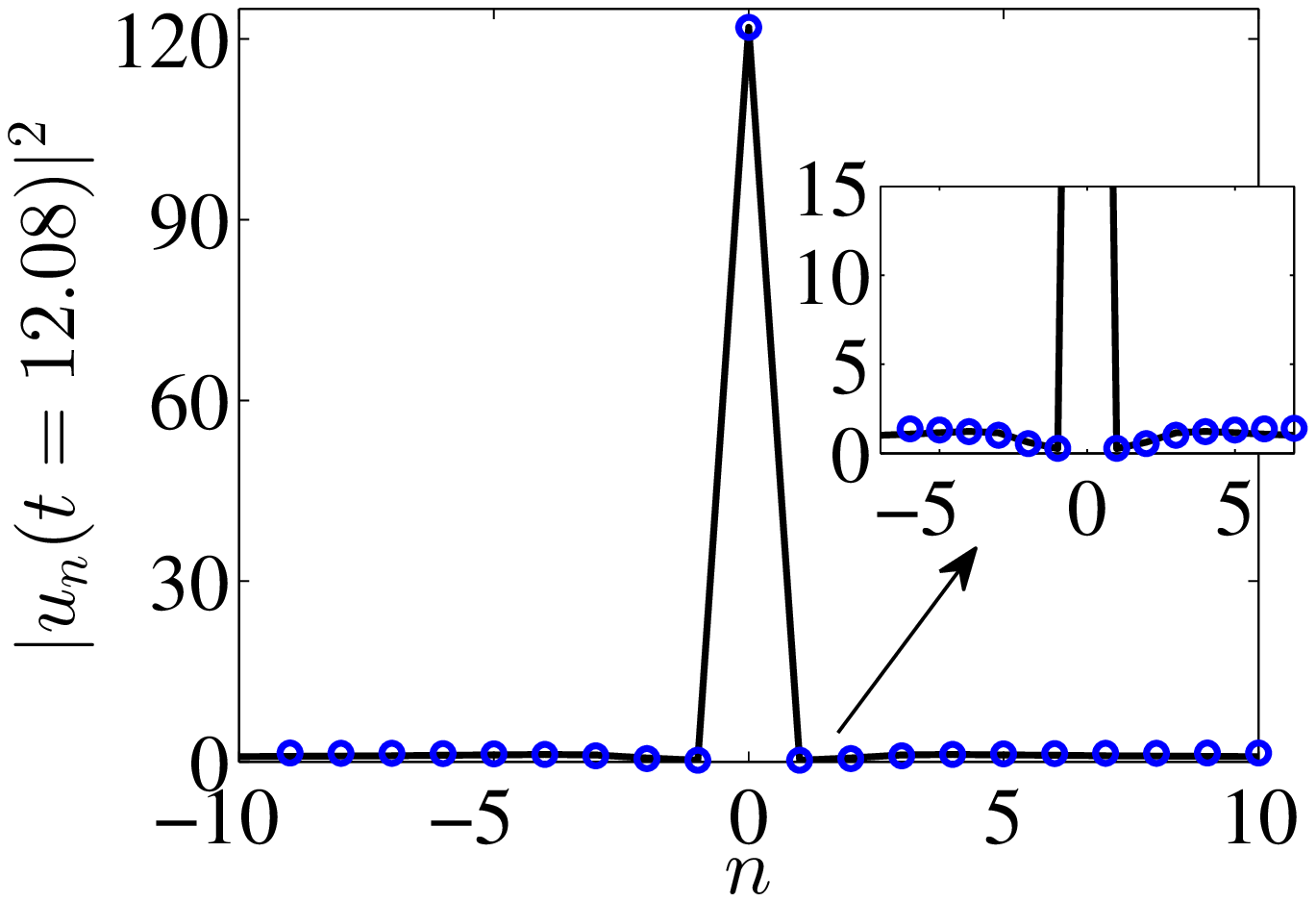}\\
\includegraphics[height=.18\textheight, angle =0]{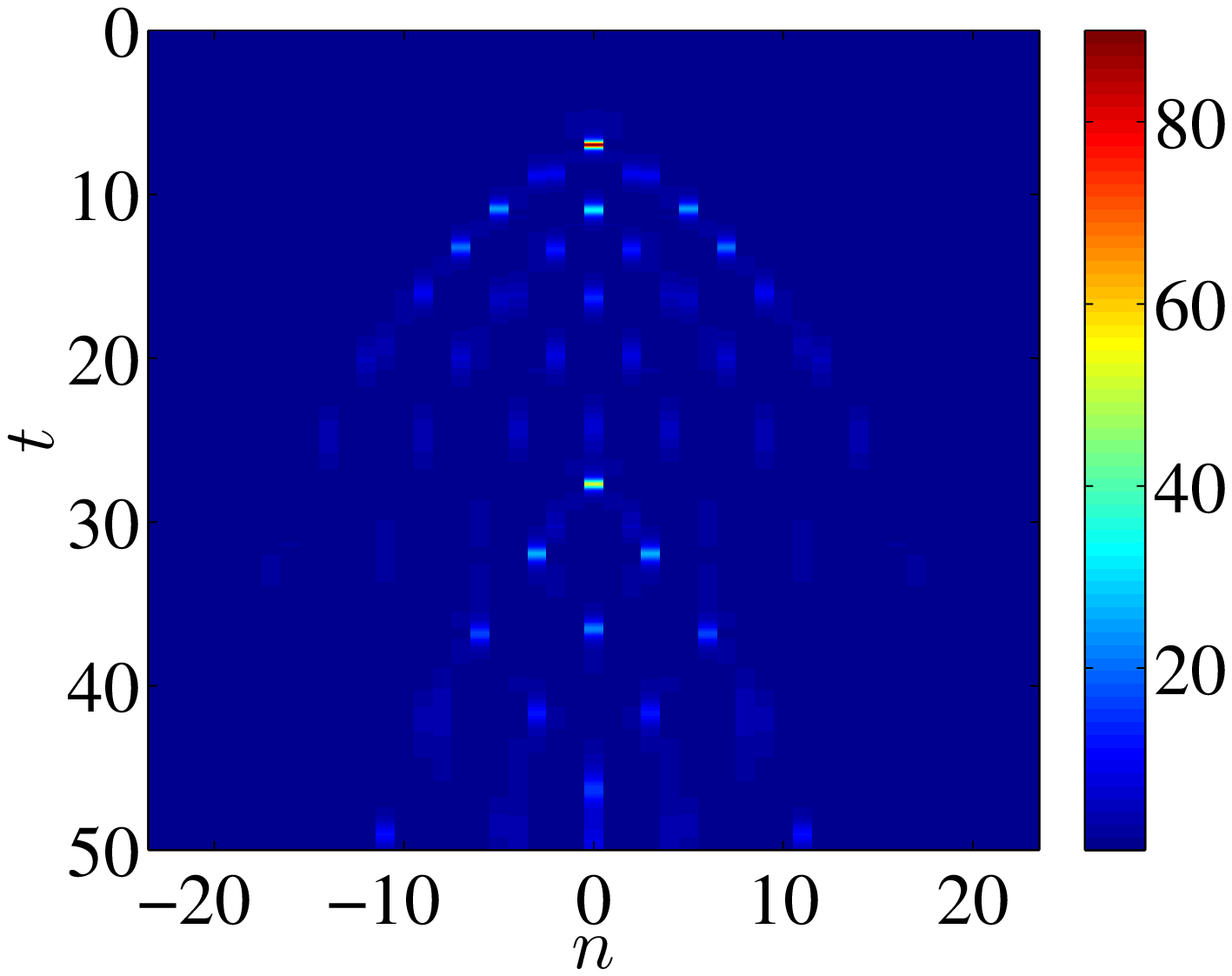}
\includegraphics[height=.18\textheight, angle =0]{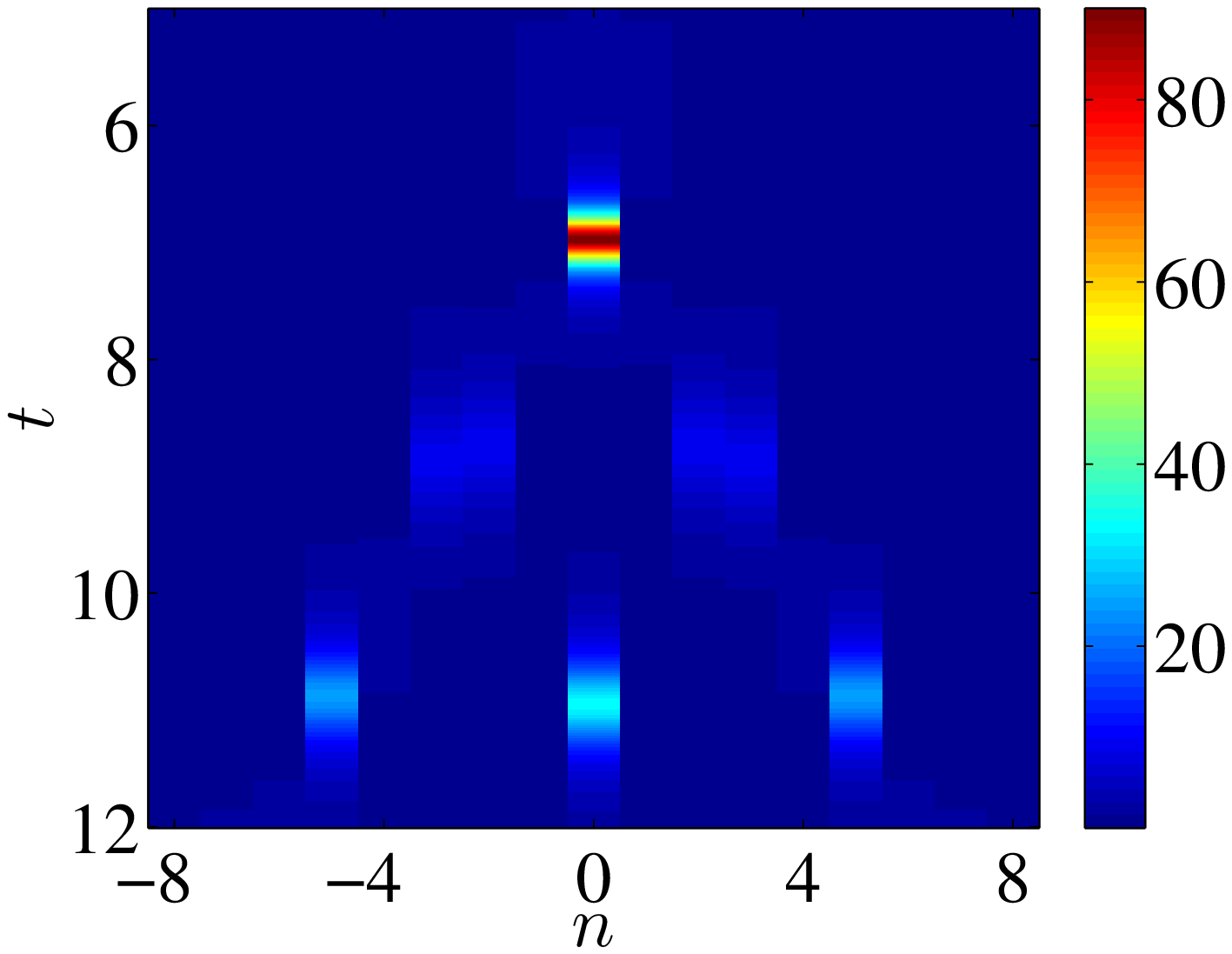}
\includegraphics[height=.18\textheight, angle =0]{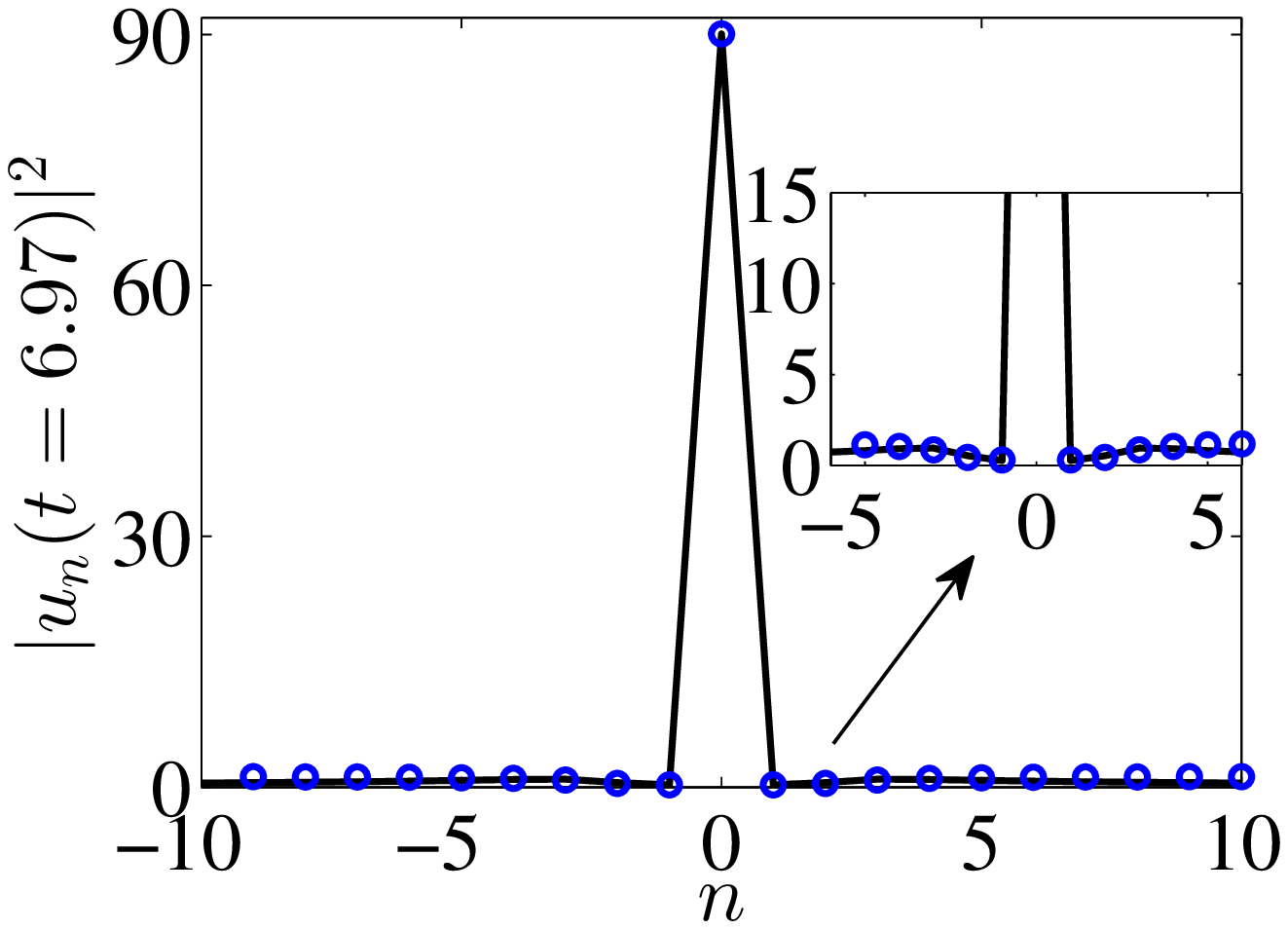}
\end{center}
\caption{
(Color online) Numerical results for the Ablowitz-Ladik (AL)
model [cf. Eq.~\eqref{gener_model} for $\mu=0$] with Gaussian 
initial data for $\sigma=30$ (top row) and $\sigma=15$ (bottom 
row). The left column shows the spatiotemporal evolution of the 
density $|u_{n}|^{2}$ while its zoom-in is presented in the middle
column for the respective cases. Spatial distribution of the 
density at $t=12.08$ and $t=6.97$ (i.e., when the first peak 
is formed) is depicted by blue open circles in the right column. 
The Peregrine solution with the same
maximal amplitude according to Eq.~\eqref{dis_per} 
is depicted by solid black lines in the right column as well.
}
\label{fig2}
\end{figure}

We now consider deviations of 
the AL model, so that we can examine whether
those features remain present as we approach the more physically
relevant DNLS limit. 
Results for a small finite value of $\mu$ (i.e., close to the AL limit), 
namely for $\mu=0.05$, are presented in the top row of Fig.~\ref{m5s33}. 
Already here, it can be seen that while the phenomenology of features
associated with a gradient catastrophe is still present, the structure
is not as closely connected to the Peregrine (it rapidly deviates from
it). This may be attributed to the fact that we only know the exact 
Peregrine soliton solution analytically in the AL case. Nevertheless, 
a more striking feature arises near $n=0$ in the top panel, and 
more drastically so throughout the evolution in the bottom panel, where 
$\mu=0.8$ (i.e., closer to the DNLS limit).
This has to do with the emergence of persistent breathing patterns, which
can be associated with discrete solitons (or discrete breathers) of the
DNLS model. These have a fundamental difference in their phenomenology
in comparison with the Peregrine in that while they are localized in
space, they are no longer localized also in time (the latter is a crucial feature
for waves that ``appear out of nowhere and disappear without a
trace''~\cite{akhmpla2}). It is important that such breathing,
persistent features emerge not only for the large $\mu$ values
of the bottom panel (where one can argue that we are rather
far away from integrability), but also even at rather small values
such as $\mu=0.05$ of the top panel. To further test this, we
have conducted simulations at smaller values of $\mu$ (down to $\mu=0.01$), confirming
that similar features do arise. In that sense, it seems more appropriate
to argue --in line also with the observations of~\cite{tsironis}--
that the integrable limit is special and, in a sense, 
seemingly rather singular as regards the persistence of 
rogue wave structures. Nevertheless, this is not a proof, and
there may well be perturbations for which the structure survives.
This topic certainly merits further investigation.

\begin{figure}[tbp]
\begin{center}
\includegraphics[height=.18\textheight, angle =0]{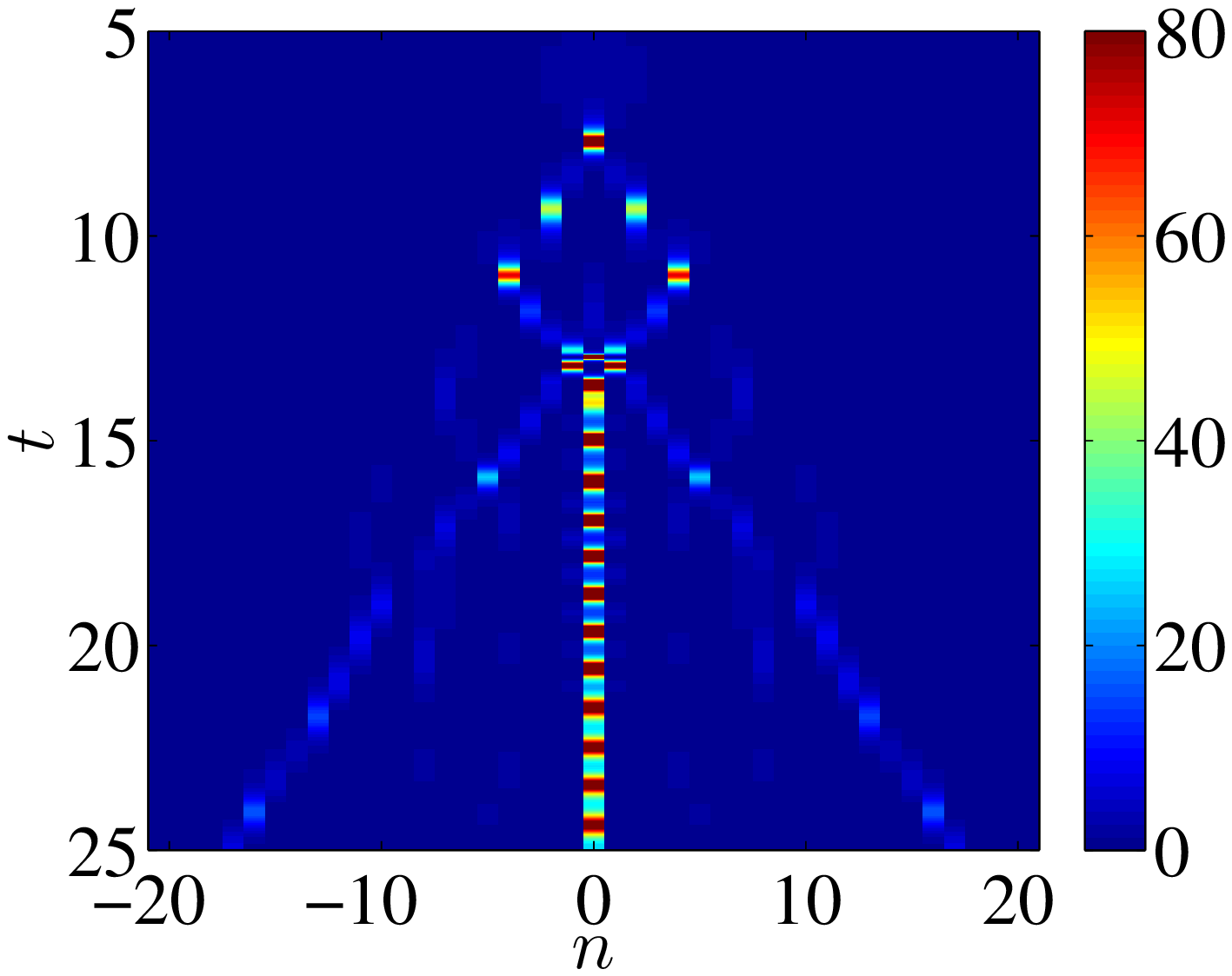}
\includegraphics[height=.18\textheight, angle =0]{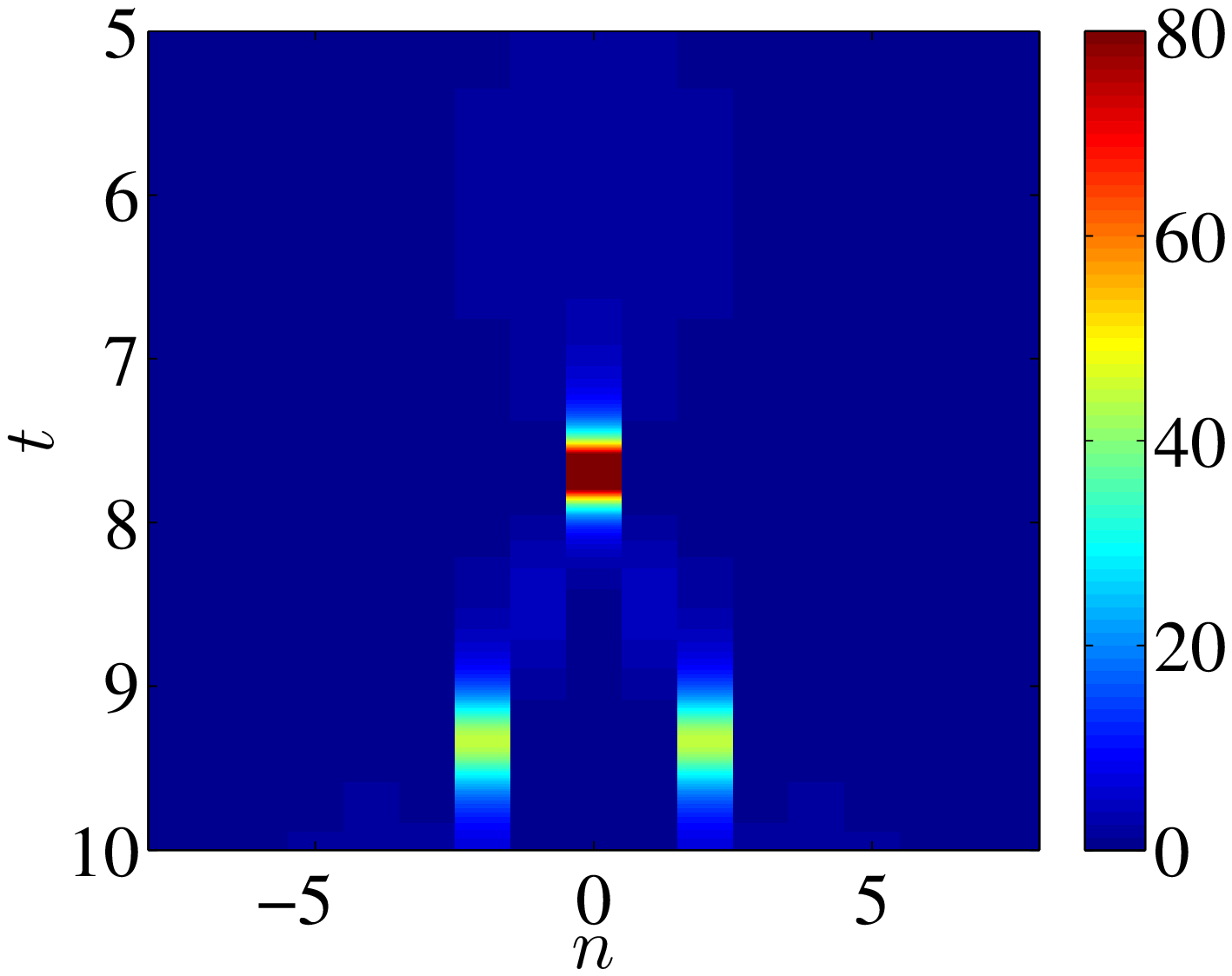}
\includegraphics[height=.18\textheight, angle =0]{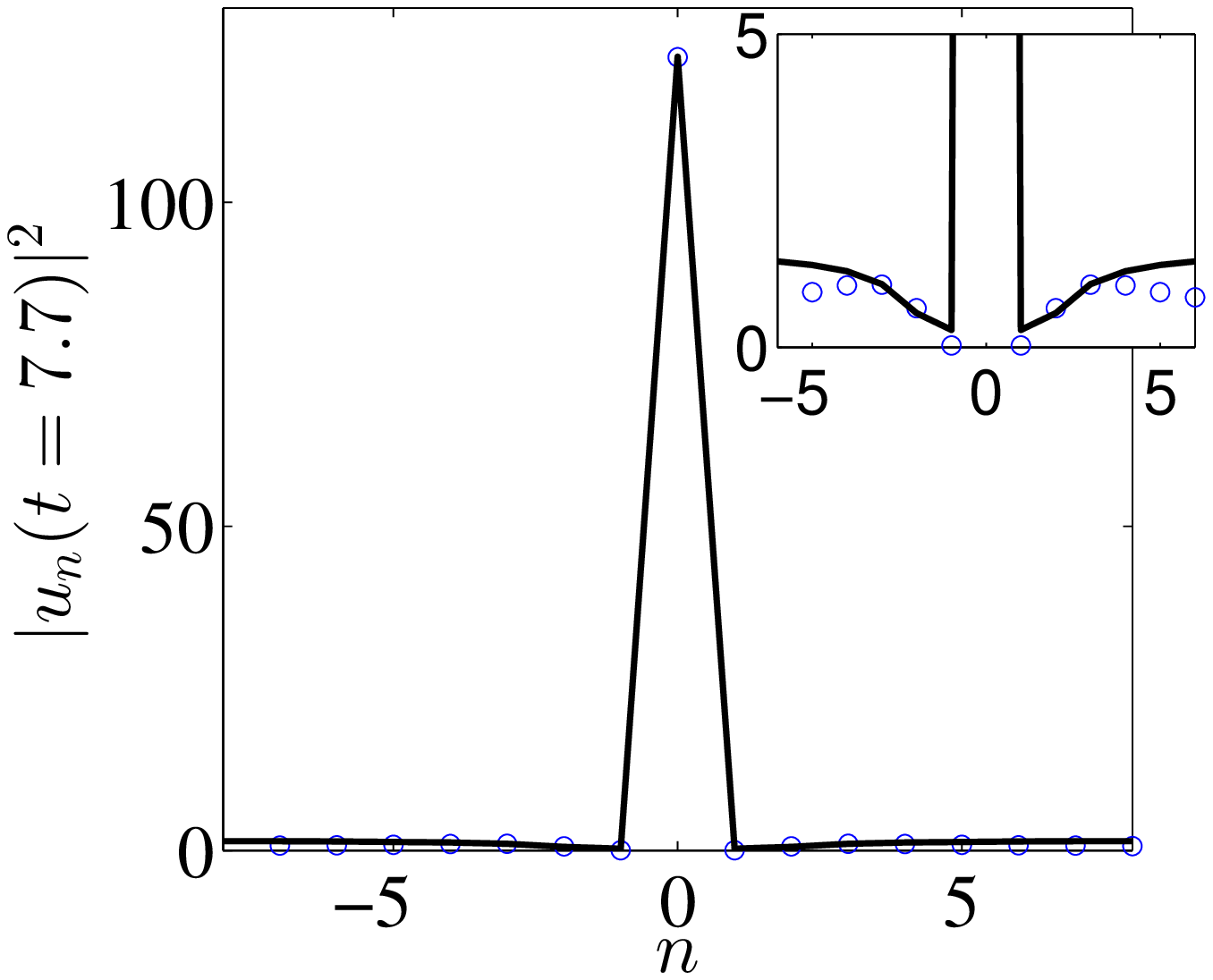}\\
\includegraphics[height=.18\textheight, angle =0]{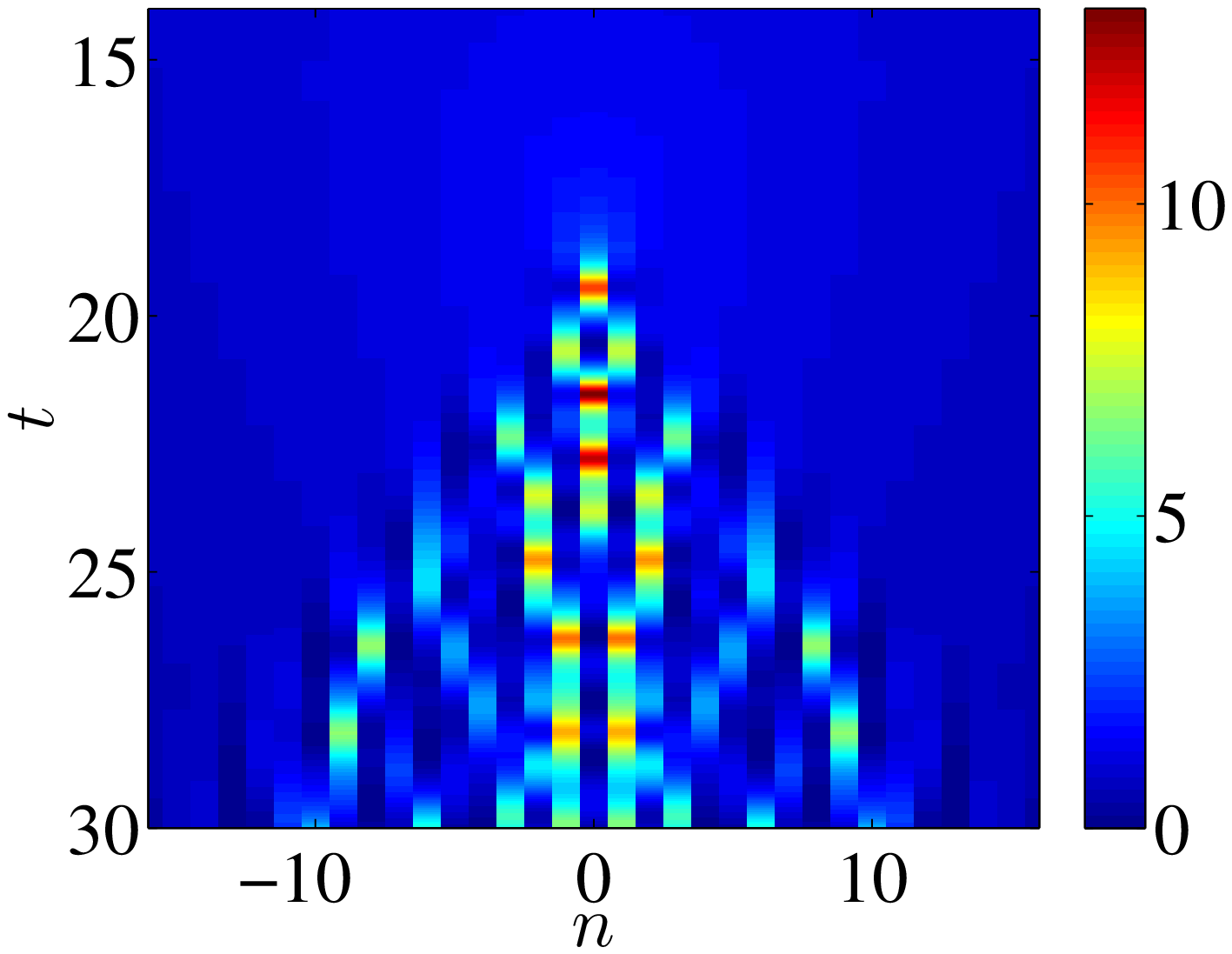}
\includegraphics[height=.18\textheight, angle =0]{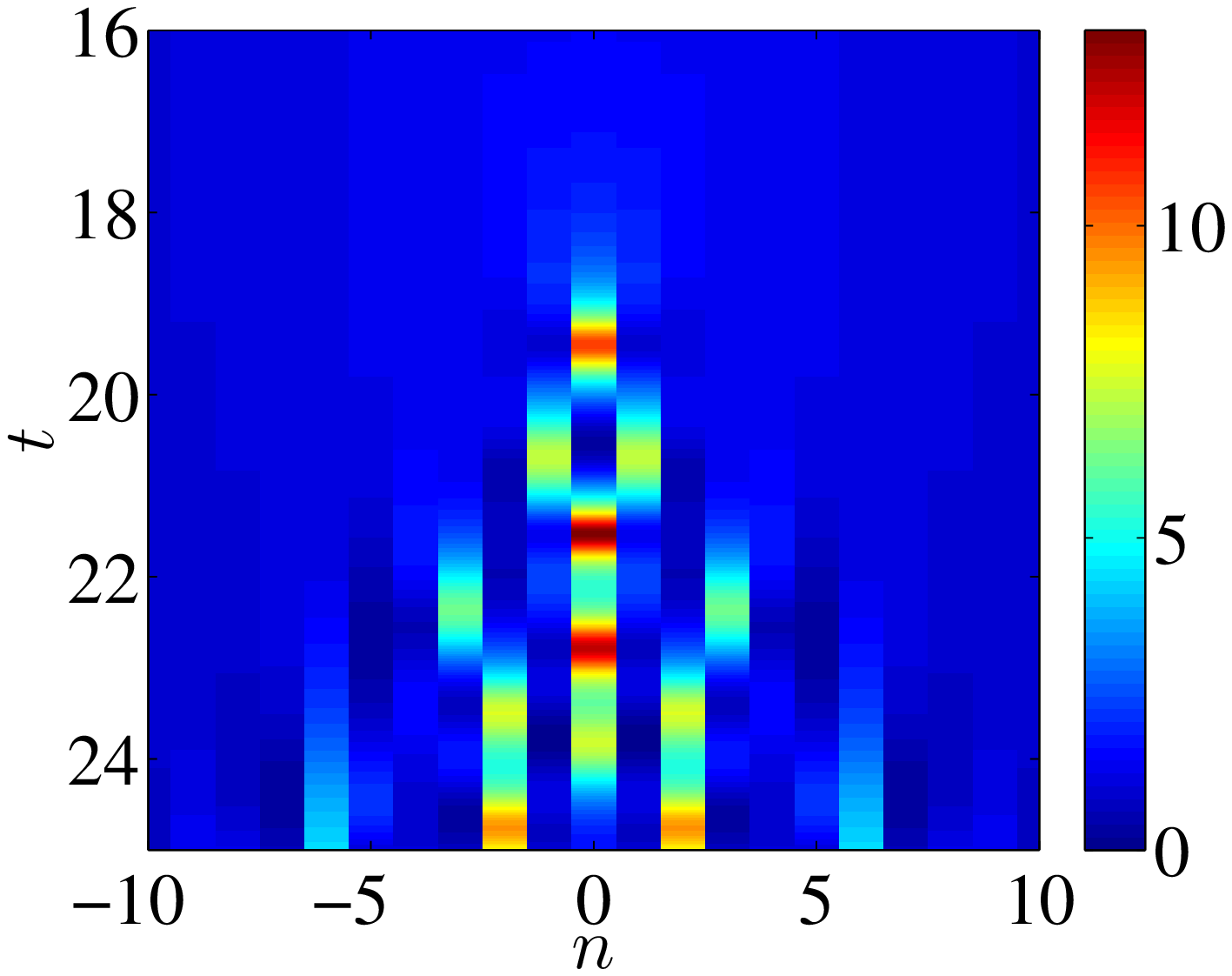}
\includegraphics[height=.18\textheight, angle =0]{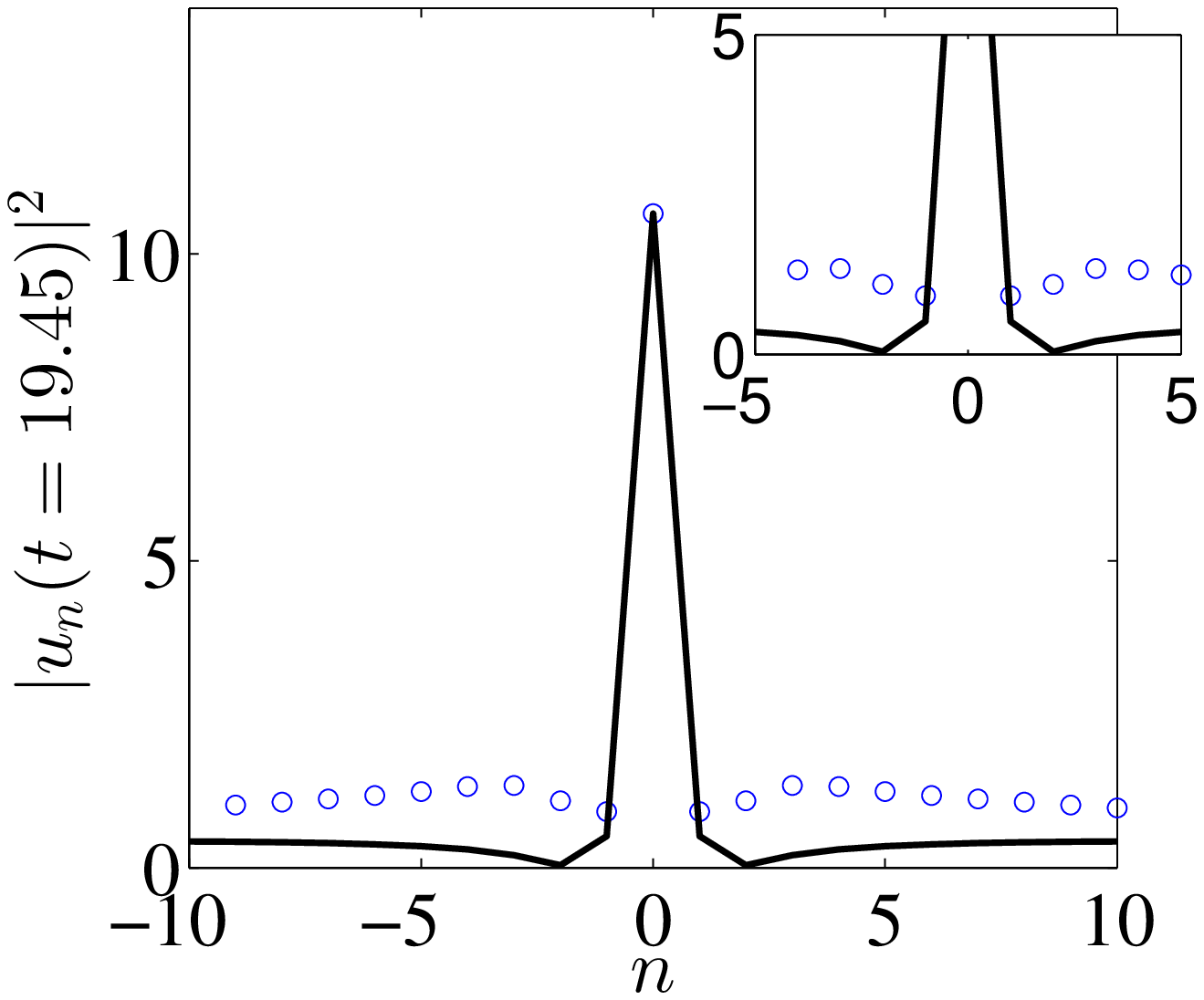}
\end{center}
\caption{
(Color online) Numerical results for the Salerno
model [cf. Eq.~\eqref{gener_model}] with Gaussian 
initial data for $\mu=0.05$ and $\sigma=3.3$ (top row) and for $\mu=0.8$ 
and $\sigma=7.4$ (bottom row). The left column shows the spatiotemporal 
evolution of the density $|u_{n}|^{2}$, while its zoom-in is presented in 
the middle column. The spatial distribution of the density at $t=7.7$ and 
$t=19.45$ (i.e., when the first peak is formed) is depicted by blue open 
circles in the right column. The best-fit Peregrine soliton solution, according to 
Eq.~\eqref{dis_per}, is depicted by solid black lines in the right column as well.
Notice the (progressively) deteriorating agreement between the two.
}
\label{m5s33}
\end{figure}

\begin{figure}[tbp]
\begin{center}
\includegraphics[height=.18\textheight, angle =0]{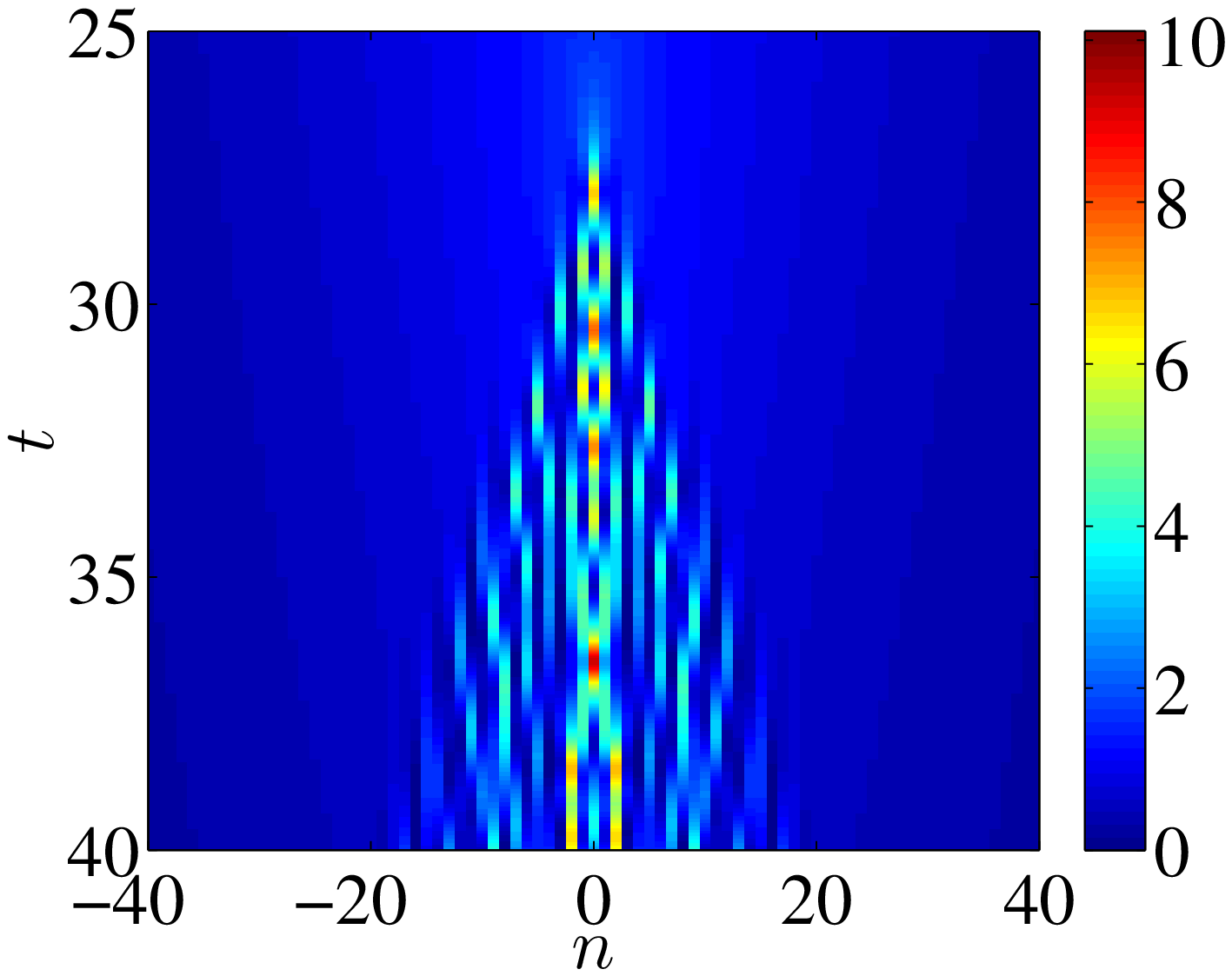}
\includegraphics[height=.18\textheight, angle =0]{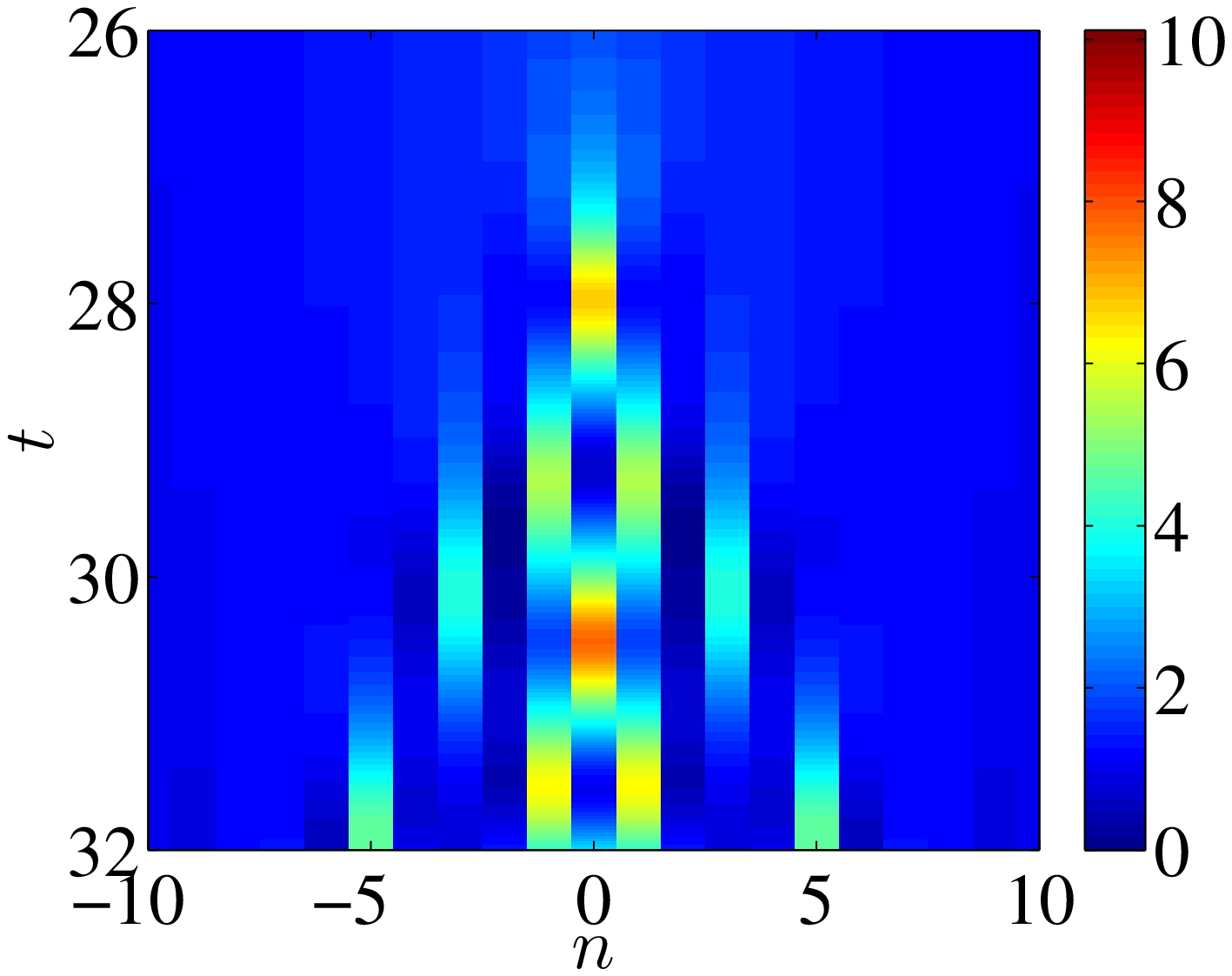}
\includegraphics[height=.18\textheight, angle =0]{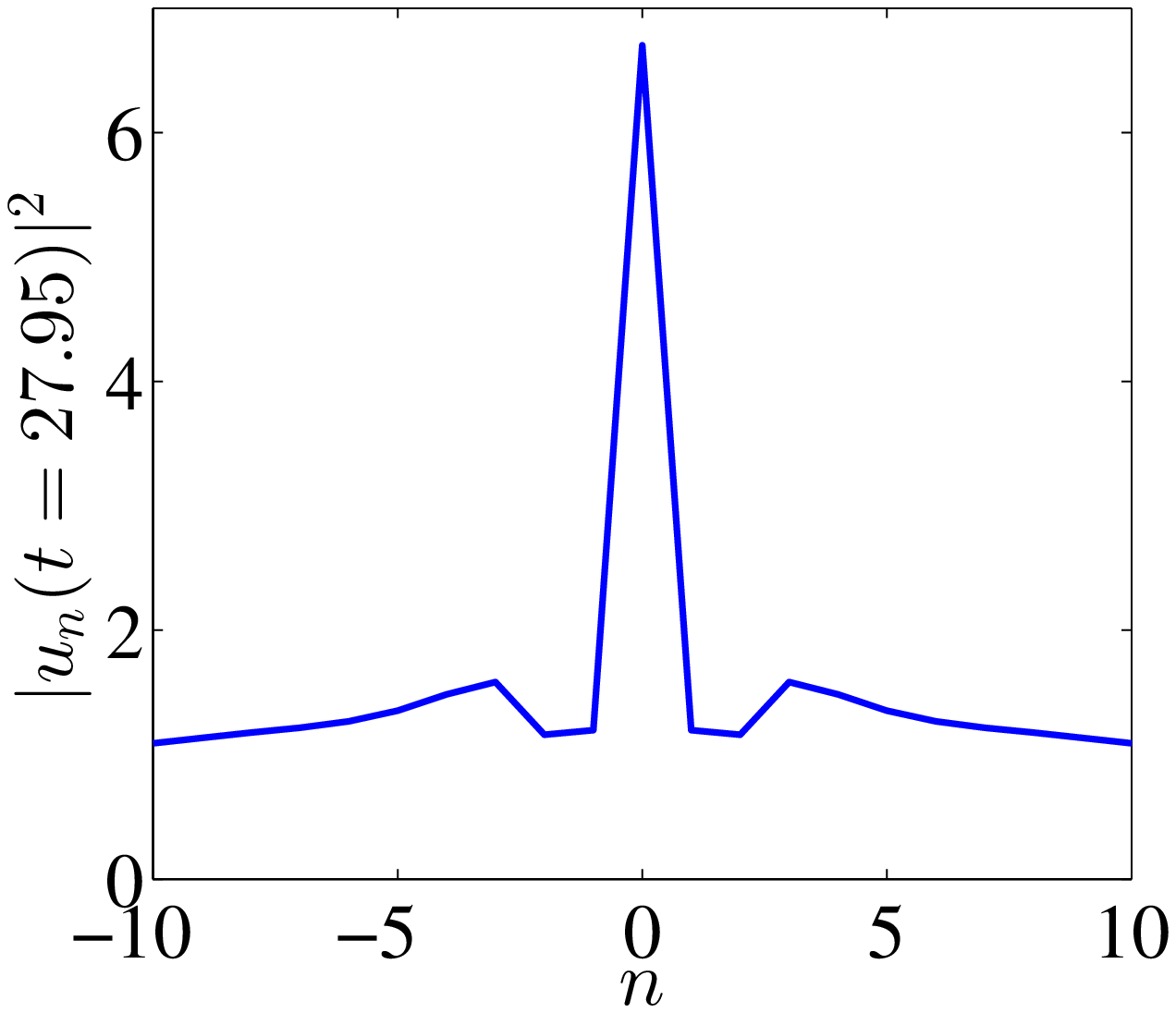}\\
\includegraphics[height=.18\textheight, angle =0]{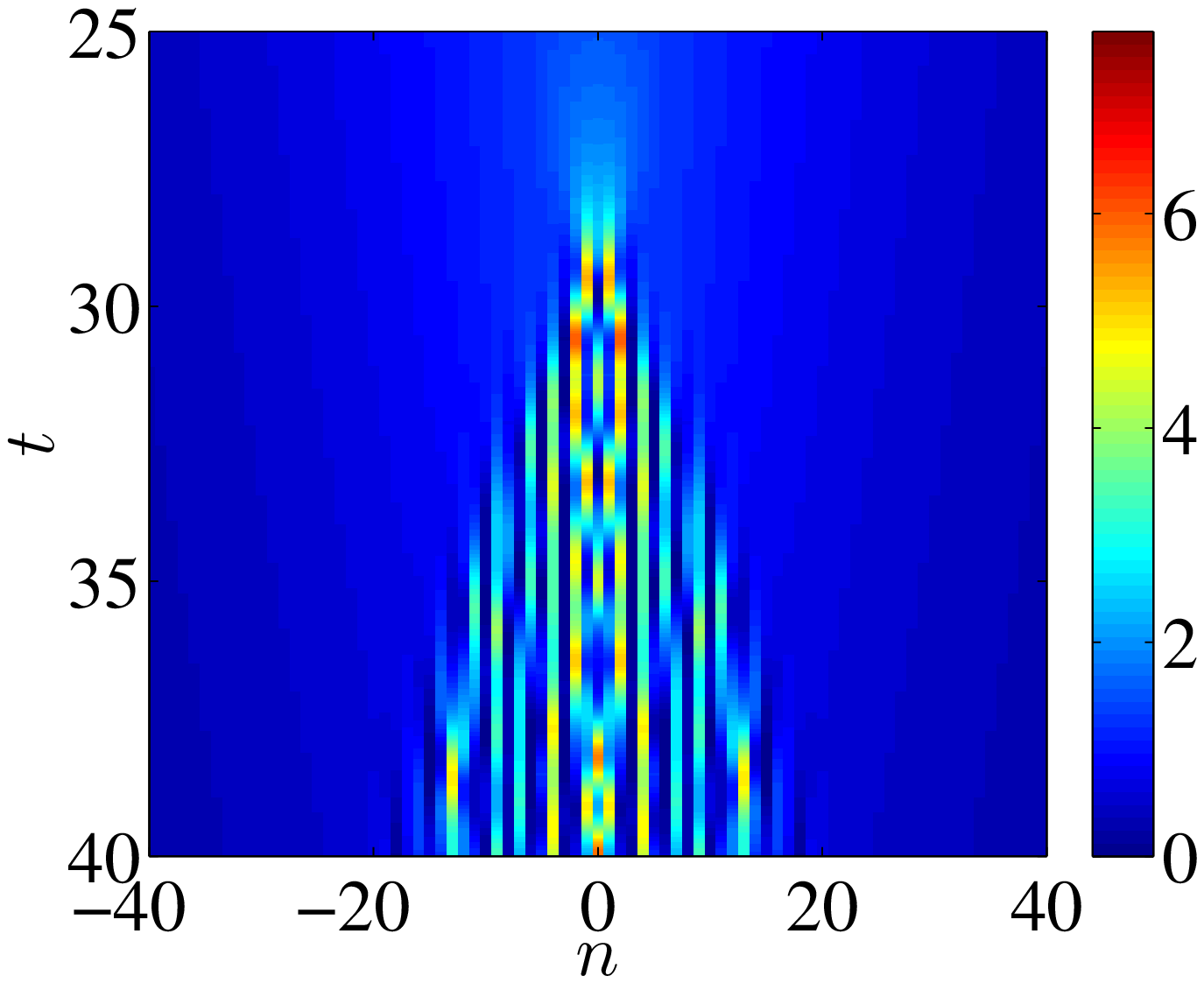}
\includegraphics[height=.18\textheight, angle =0]{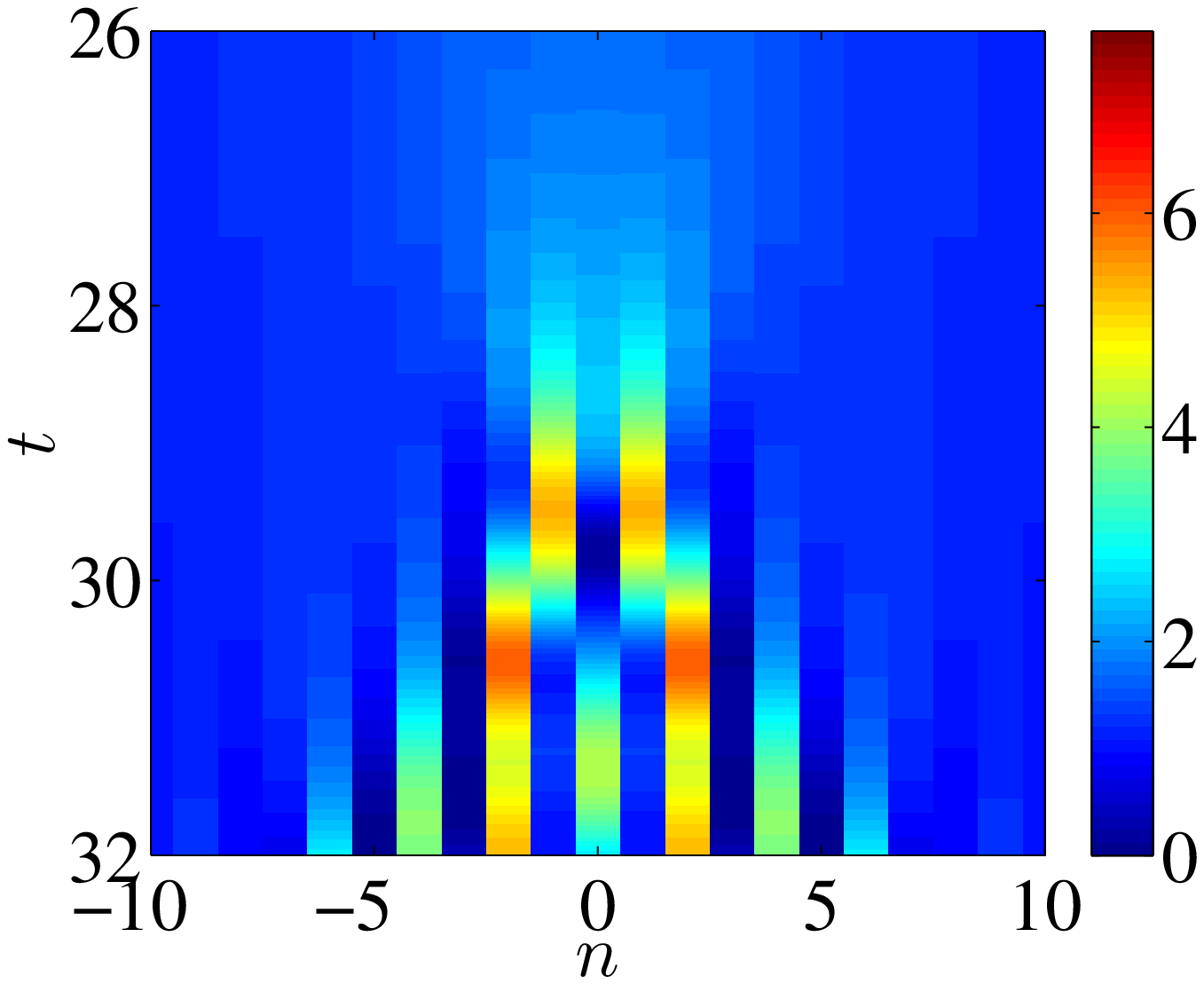}
\includegraphics[height=.18\textheight, angle =0]{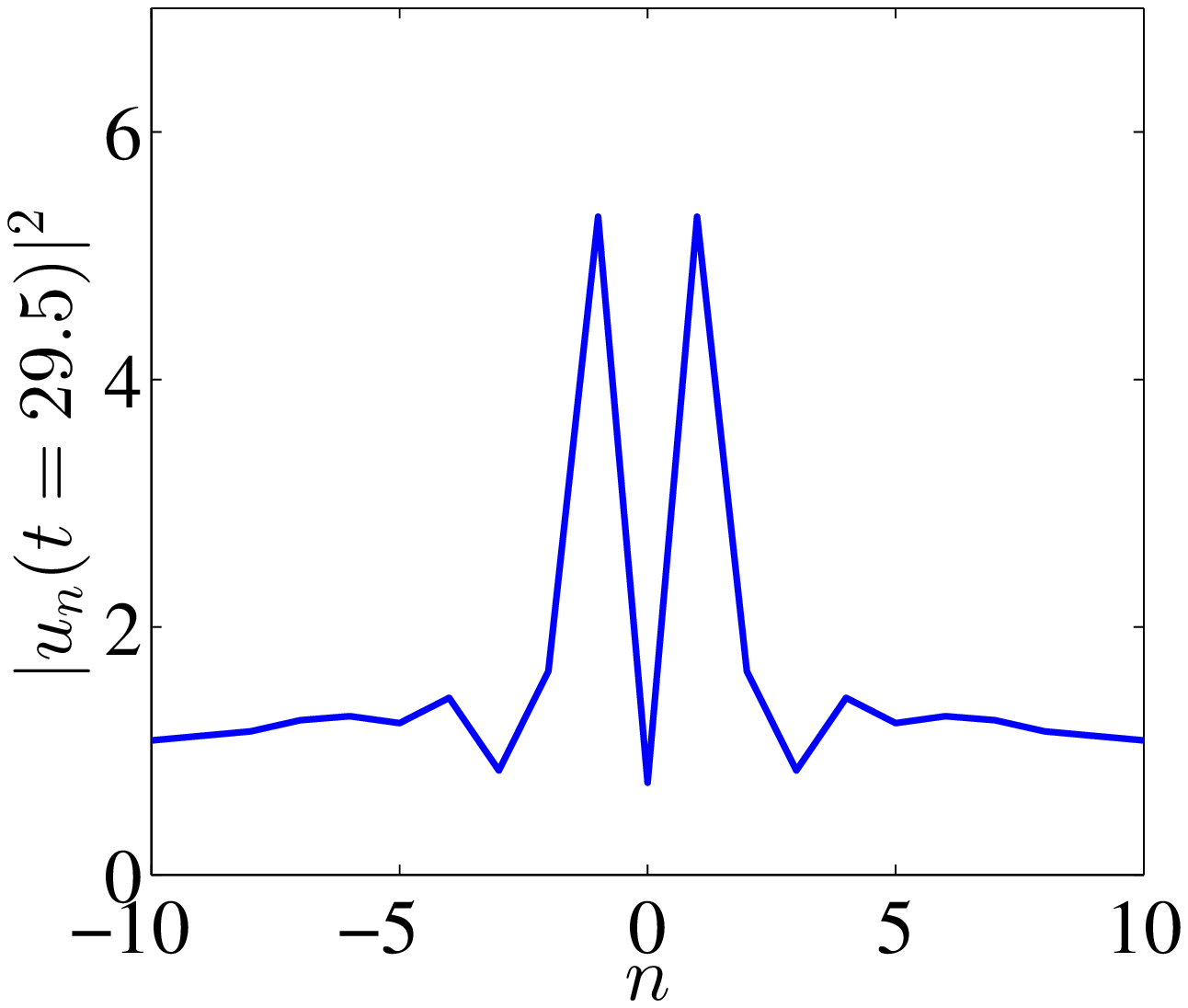}\\
\includegraphics[height=.18\textheight, angle =0]{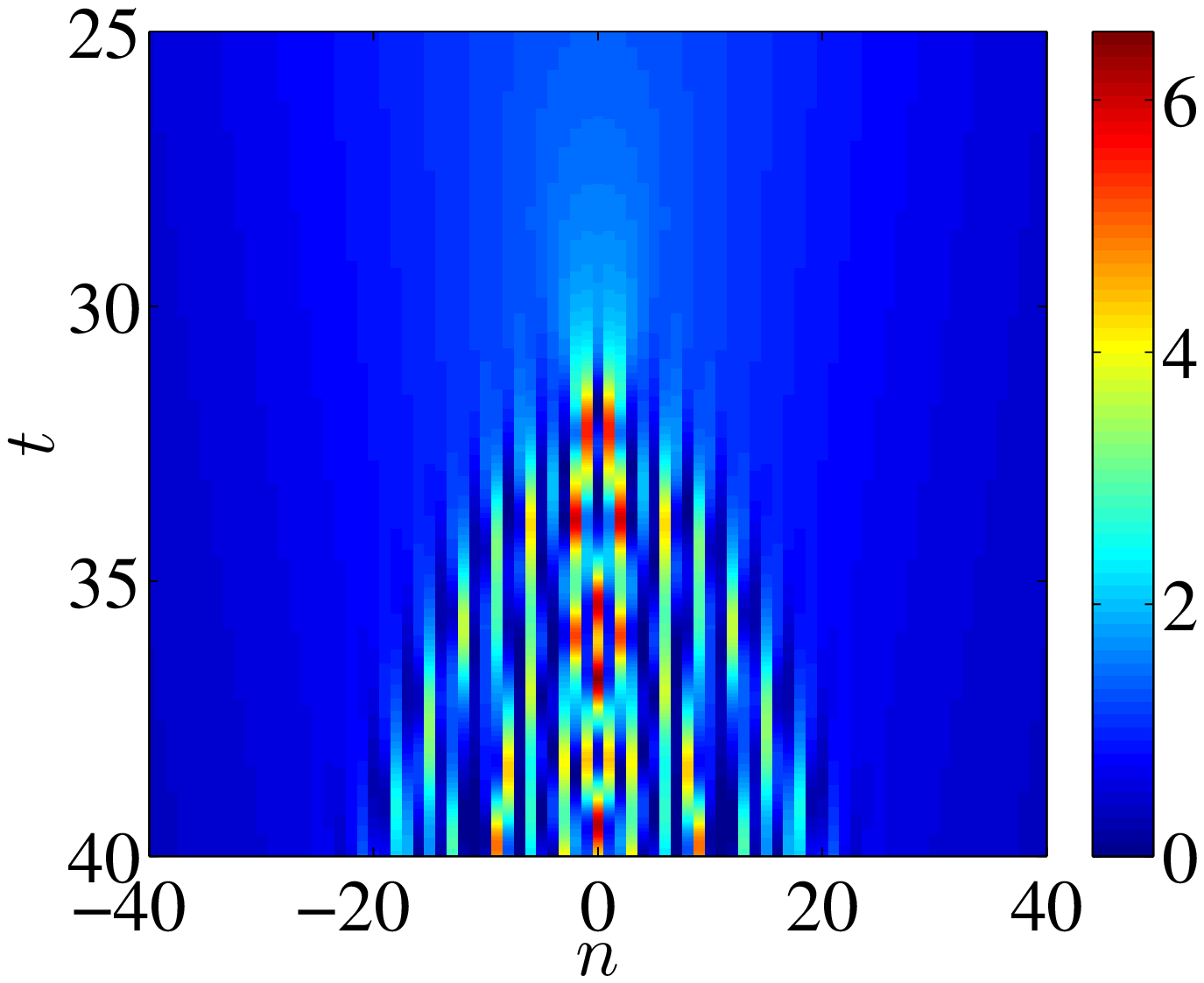}
\includegraphics[height=.18\textheight, angle =0]{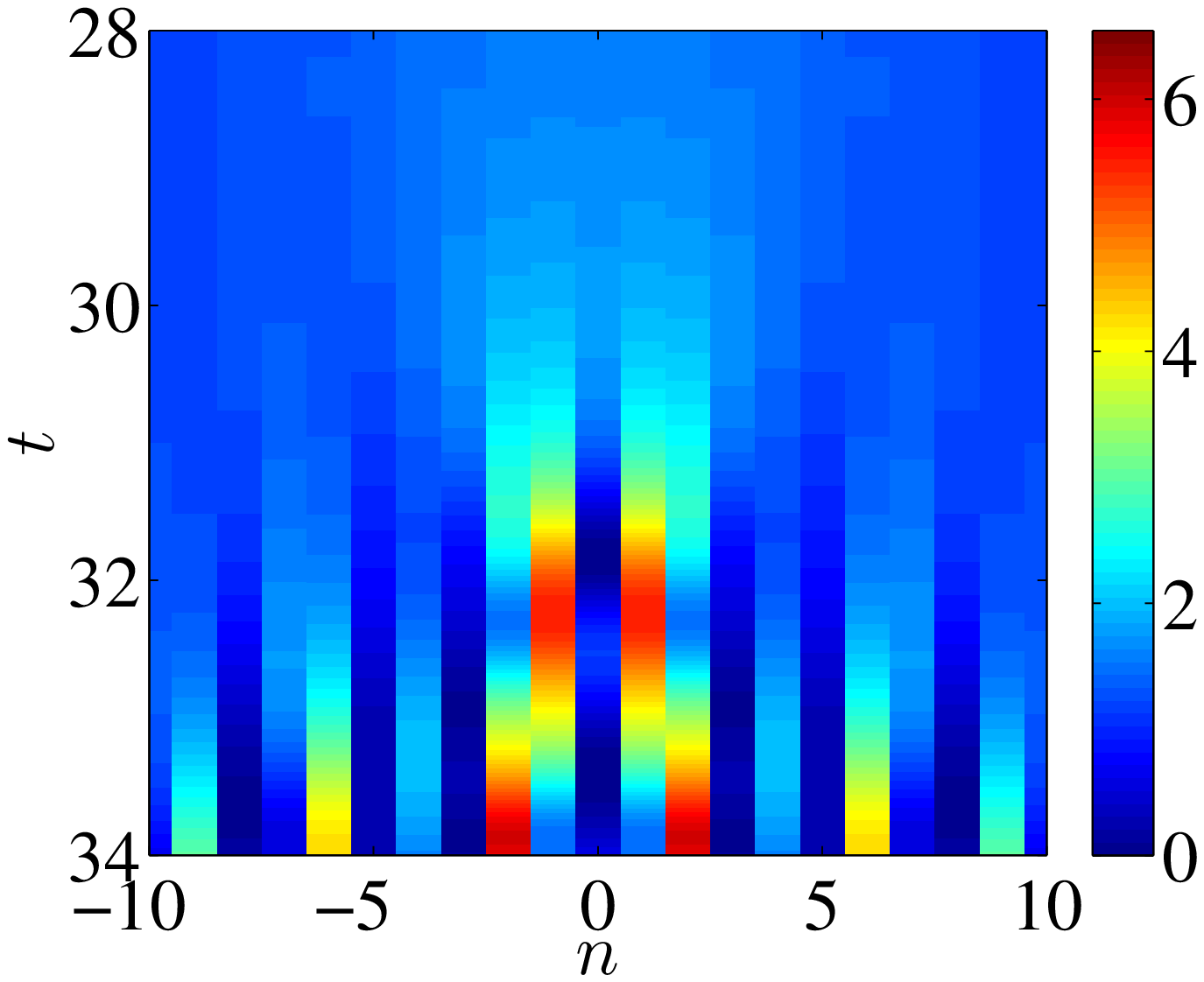}
\includegraphics[height=.18\textheight, angle =0]{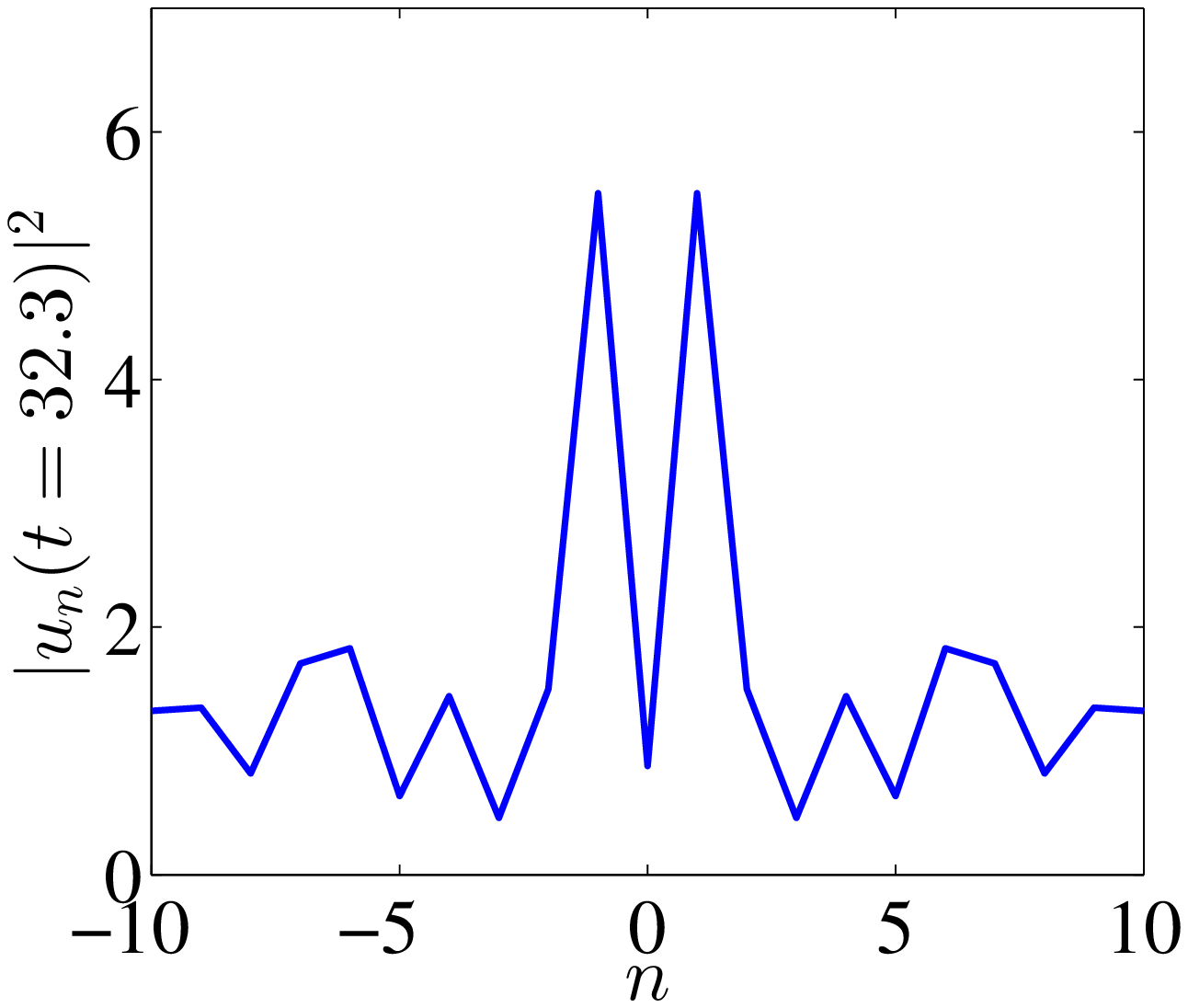}
\end{center}
\caption{
(Color online) Numerical results for the Salerno 
model [cf. Eq.~\eqref{gener_model}] in the DNLS limit, i.e., for $\mu=1$, with Gaussian 
initial data, 
with $\sigma=9.6$ (top row), $\sigma=9.8$ 
(center row) and  $\sigma=11$ (bottom row). The left column shows the 
spatiotemporal evolution of the density $|u_{n}|^{2}$ while its 
zoom-in is presented in the middle column. Spatial distribution of 
the density at $t=27.95$ (top), $t=29.5$ (center) and $t=32.3$ (bottom) 
(i.e., when the first peak is formed) is depicted by solid blue lines
in the right column.
}
\label{m95}
\end{figure}
%

We now consider 
the other end/limit, and study 
the special DNLS case of $\mu=1$ for different values of $\sigma$; 
pertinent results are shown in Fig.~\ref{m95}.
Here, too, we can observe that the focusing process may have
a gradient catastrophe character --see especially the top
right panel of the figure-- although the phenomenon
seems less extreme in the middle and bottom right panels.
Once again, we furthermore emphasize the
apparent persistence of the structures that are emerging
(notice the ``threads'' of the relevant patterns and their
``breathing'' in the left and middle panels). This implies
that the relevant configurations are less structurally proximal
to Peregrine solitons and more so to discrete solitonic structures.


%
%

In a broader picture, the panels displayed in Fig.~\ref{c_panel} 
provide an overview of those features. This figure is intended as an effective
``two-parameter diagram'' from the point of view of the dynamics of the
Salerno model for 
initial conditions of different width, and different 
values of the parameter $\mu$, ranging from the integrable 
to progressively more non-integrable cases 
(an animated figure showing the behaviour for $\sigma \in[0.1,30]$  is available at \footnote{\url{https://youtu.be/-X2XuU4o9RA}}).
The left column shows the integrable AL model as a reference. 
The middle column highlights a weak perturbation of the integrable limit for
the case of $\mu = 0.02$ (available at  \footnote{\url{https://youtu.be/BvAg-H31k6o}}). 
Finally, a substantial departure from the
integrable limit is displayed for $\mu = 0.5$ in the right column
(available at  \footnote{\url{https://youtu.be/WLW5uBVDCQI}}).
All those are compared for specific values of $\sigma$ in each row.
In the topmost row, the case of small $\sigma$, 
i.e., $\sigma = 0.6$, illustrates
a fairly similar behavior, irrespectively of the value of $\mu$.
A breathing dynamics appears to ensue (reminiscent of 2-soliton
solutions; see the relevant discussion in Ref.~\cite{stathis}) and the only
feature changing across the columns is the period of the relevant pattern.

The differences across the columns are starting to be more pronounced
in the second row for the case of $\sigma = 1.2$. The left panel,
as expected for the integrable scenario, develops a solitonic character
of a multi-soliton state (cf. also the corresponding discussion
of Ref.~\cite{stathis}). However, already at small departures from integrability
a robust structure emerges near the center and is even more pronounced
and permanent in the case of $\mu=0.5$.
These features are even further amplified in the third row
featuring the case of $\sigma = 4.2$. Here the AL pattern clearly
re-creates the Christmas-tree-type pattern that was observed in
the corresponding continuum NLS model. Nevertheless, the
non-integrable model even for a slight deviation, such as $\mu=0.02$, 
changes the dynamics drastically and no longer enables the
propagation of the dendritic structure. Instead, three filamentary 
breathing patterns appear to occur and survive as 
long-lived entities in the corresponding dynamics. This departure 
remains pronounced in the case of $\mu=0.5$ in the right panel 
of the third row. 

For even larger values of $\sigma$ the tree structure, 
which started out in a triangular shape for the AL model
(with the angle of the outer slope changing depending on $\sigma$), 
eventually changes to a parabolic shape with the extreme events packed closely together.
A rather similar behaviour is shown for $\mu = 0.02$,
while $\mu = 0.5$ is much more extreme, with some events
apparently being able to linger on for quite some time.

\begin{figure}[tbp]
\begin{center}
\includegraphics[height=.18\textheight, angle =0]{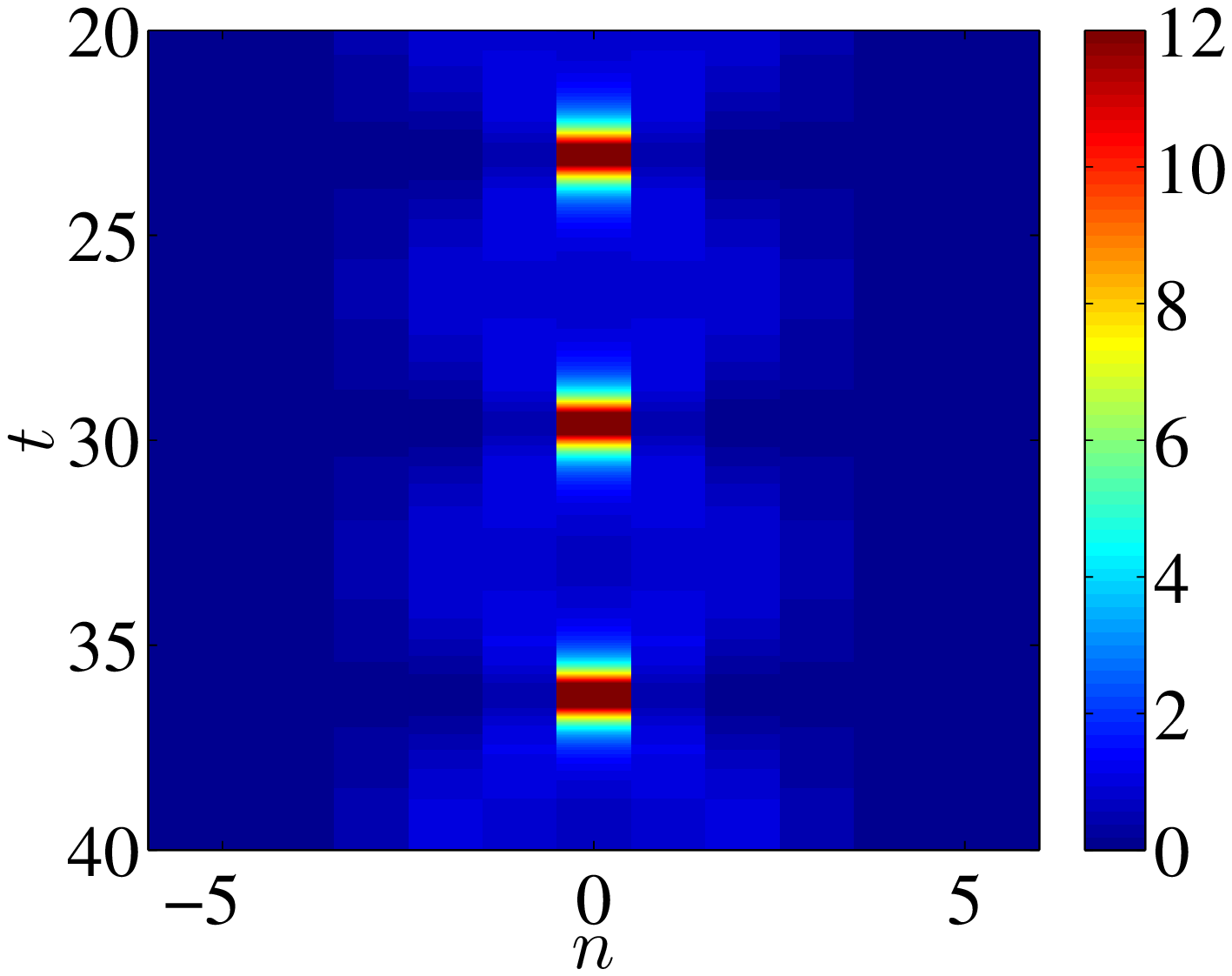}
\includegraphics[height=.18\textheight, angle =0]{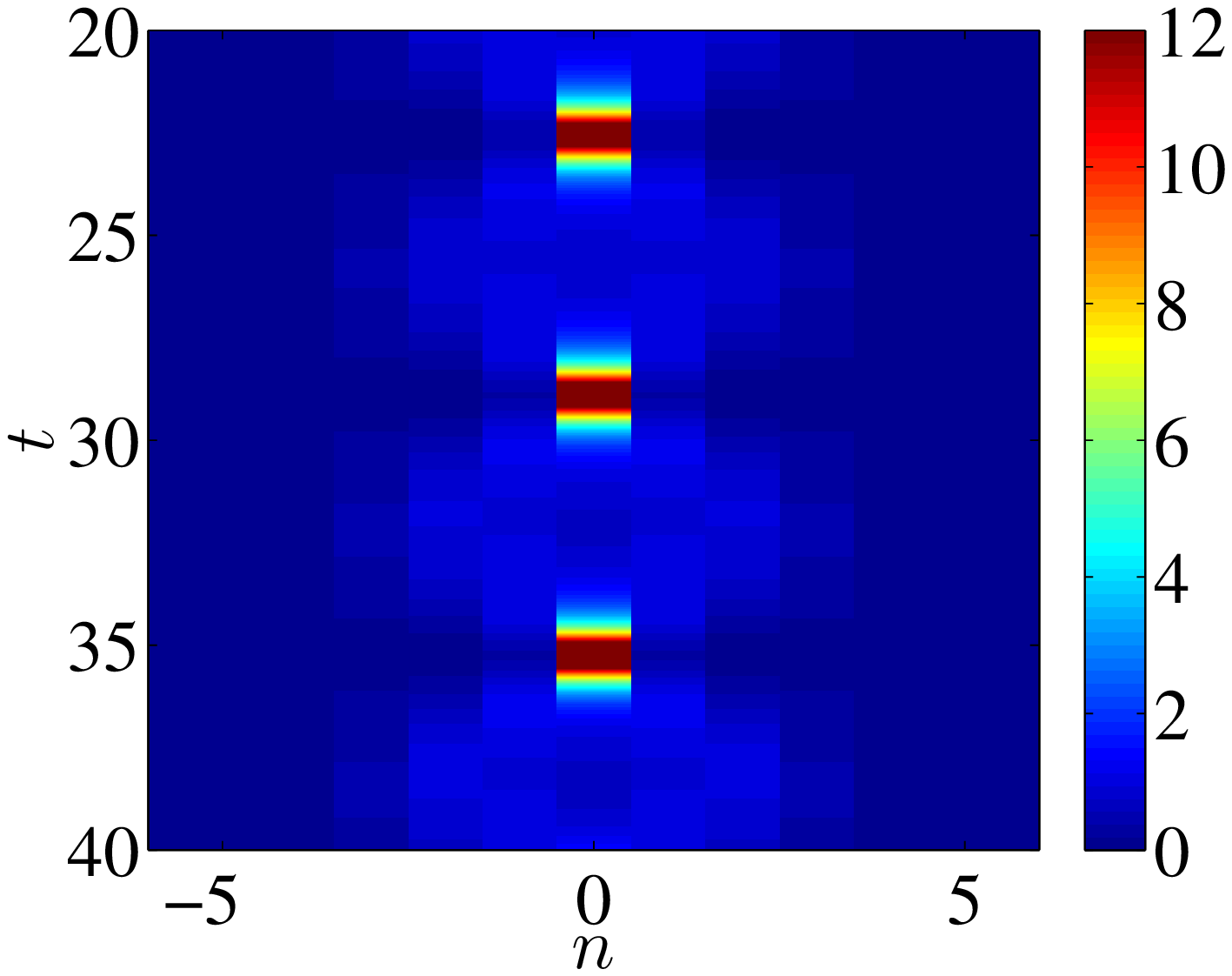}
\includegraphics[height=.18\textheight, angle =0]{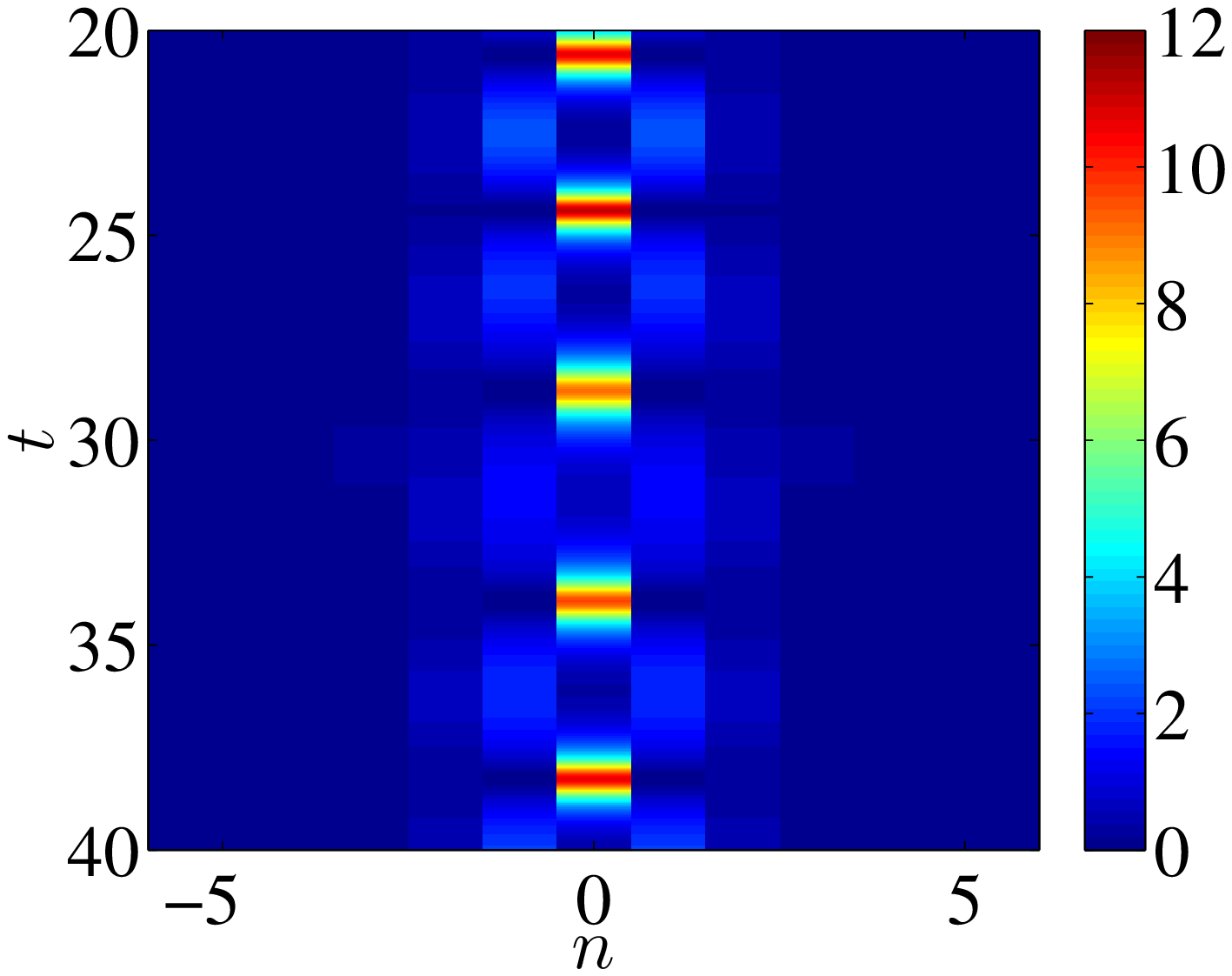}\\
\includegraphics[height=.18\textheight, angle =0]{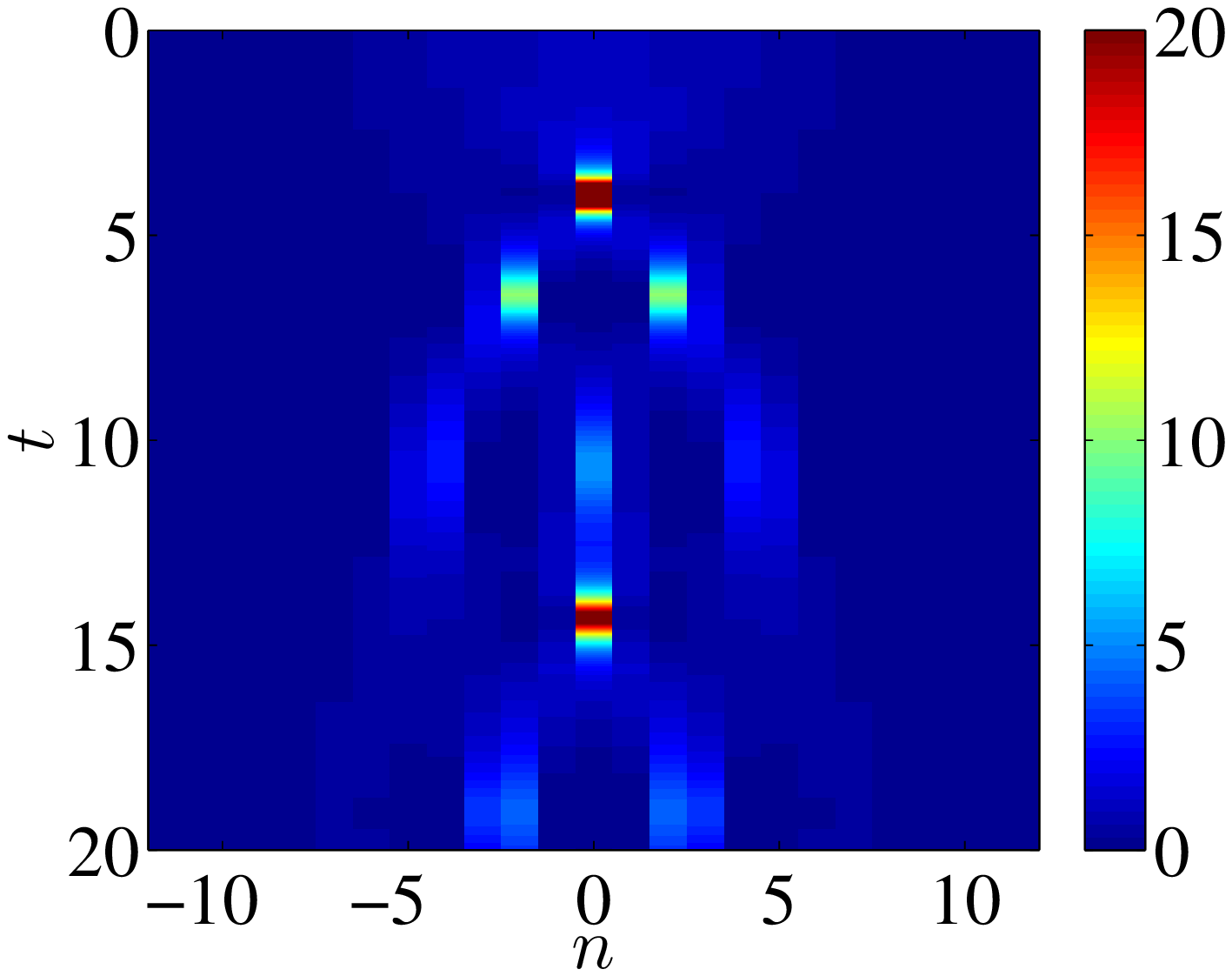}
\includegraphics[height=.18\textheight, angle =0]{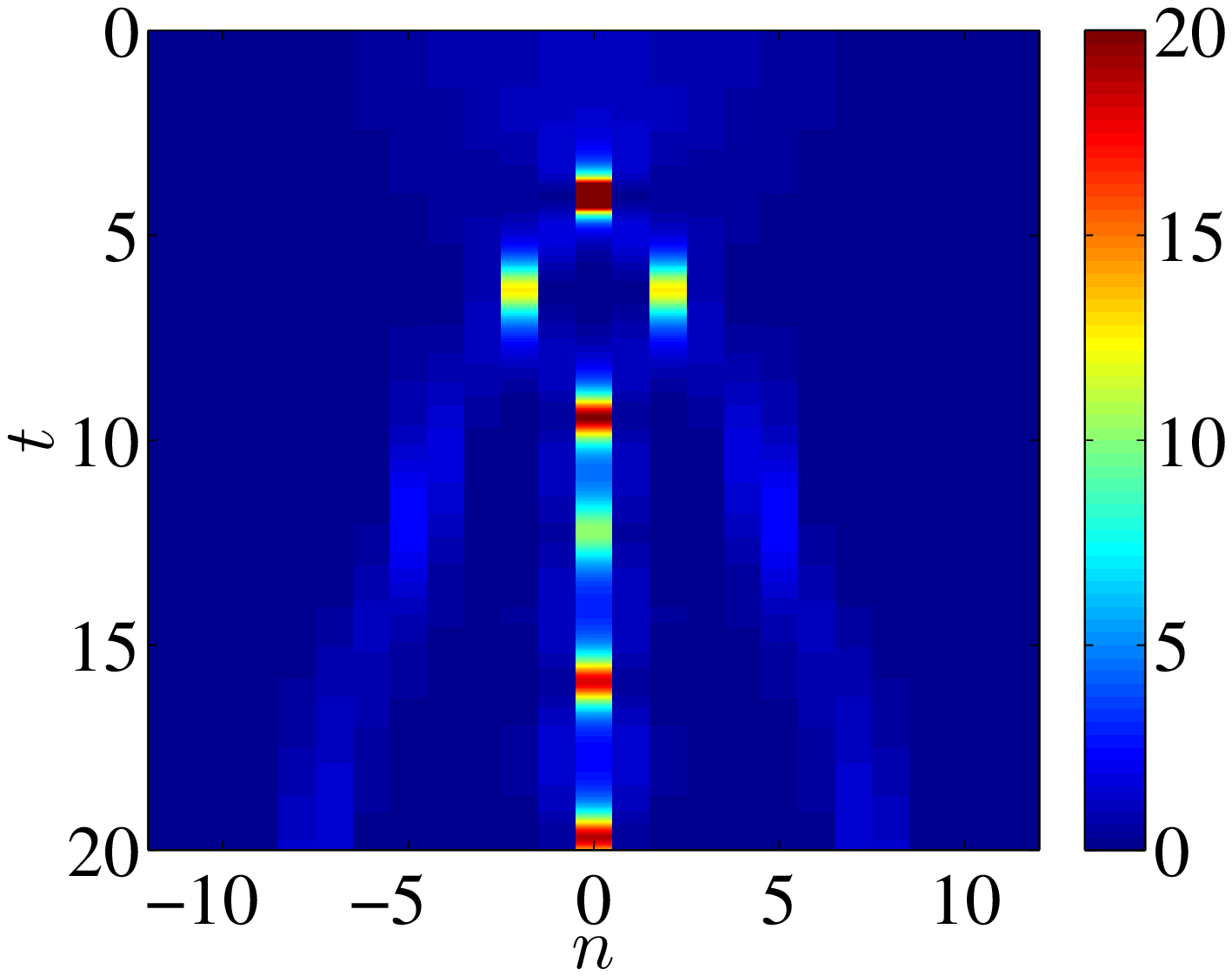}
\includegraphics[height=.18\textheight, angle =0]{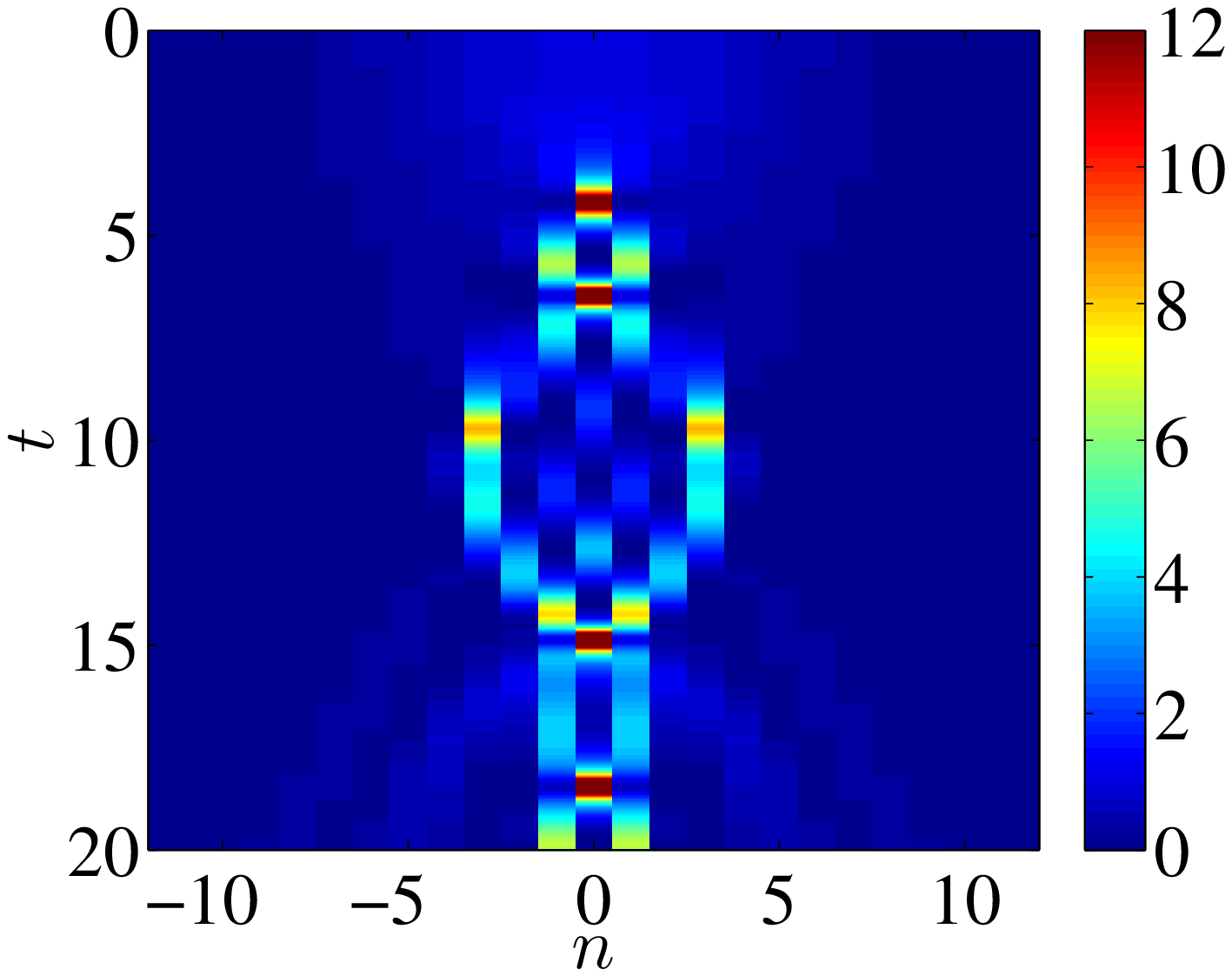}\\
\includegraphics[height=.18\textheight, angle =0]{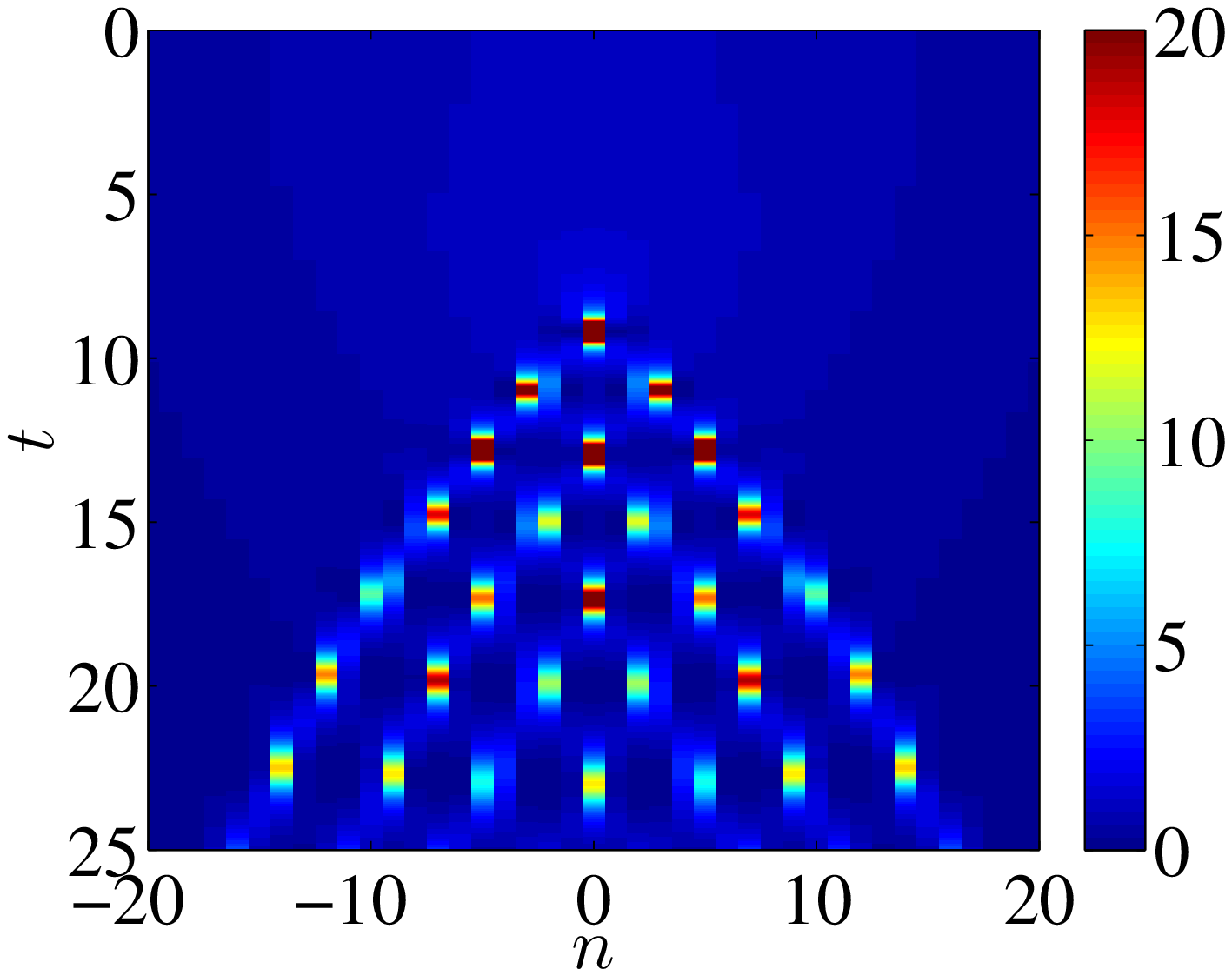}
\includegraphics[height=.18\textheight, angle =0]{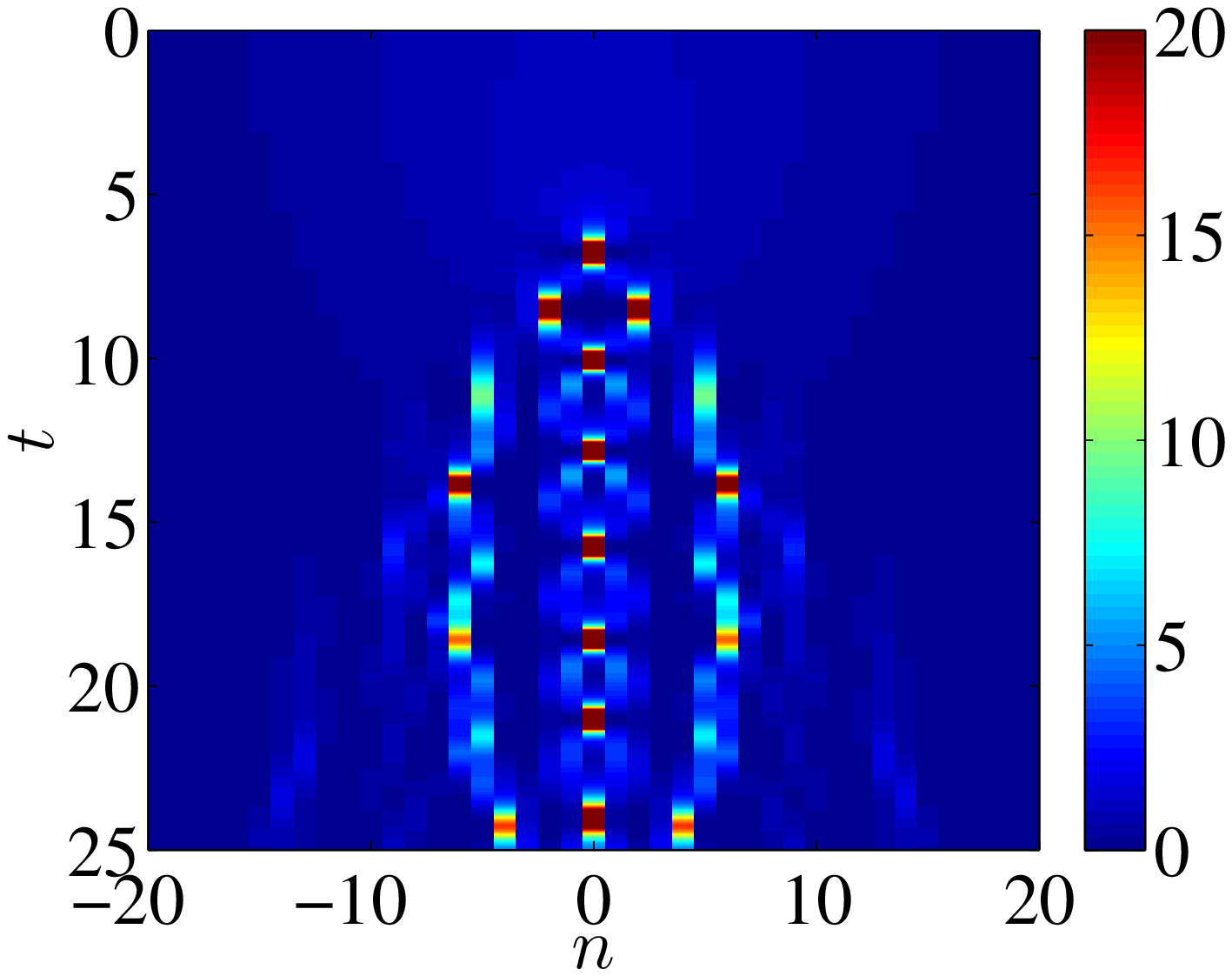}
\includegraphics[height=.18\textheight, angle =0]{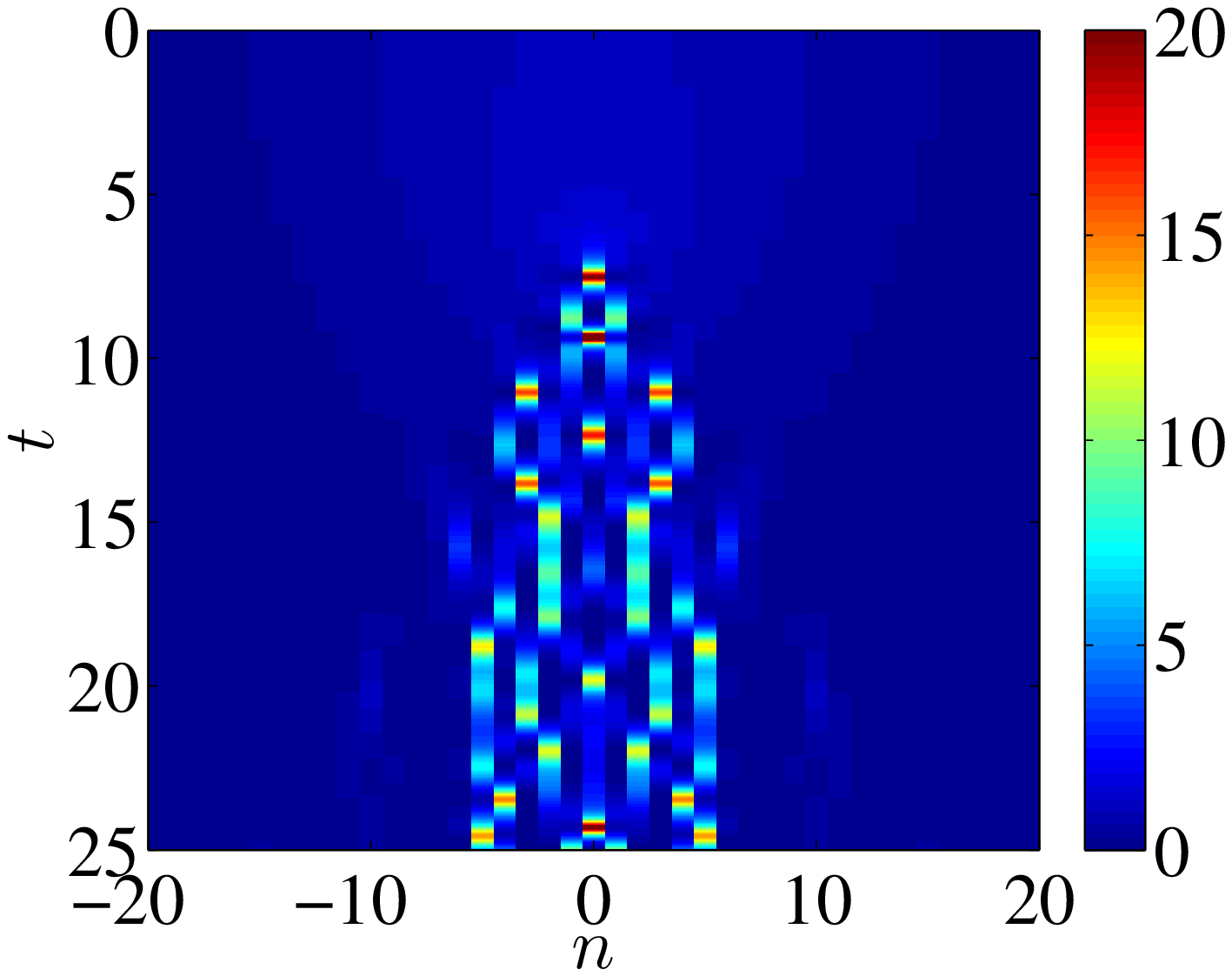}\\
\includegraphics[height=.18\textheight, angle =0]{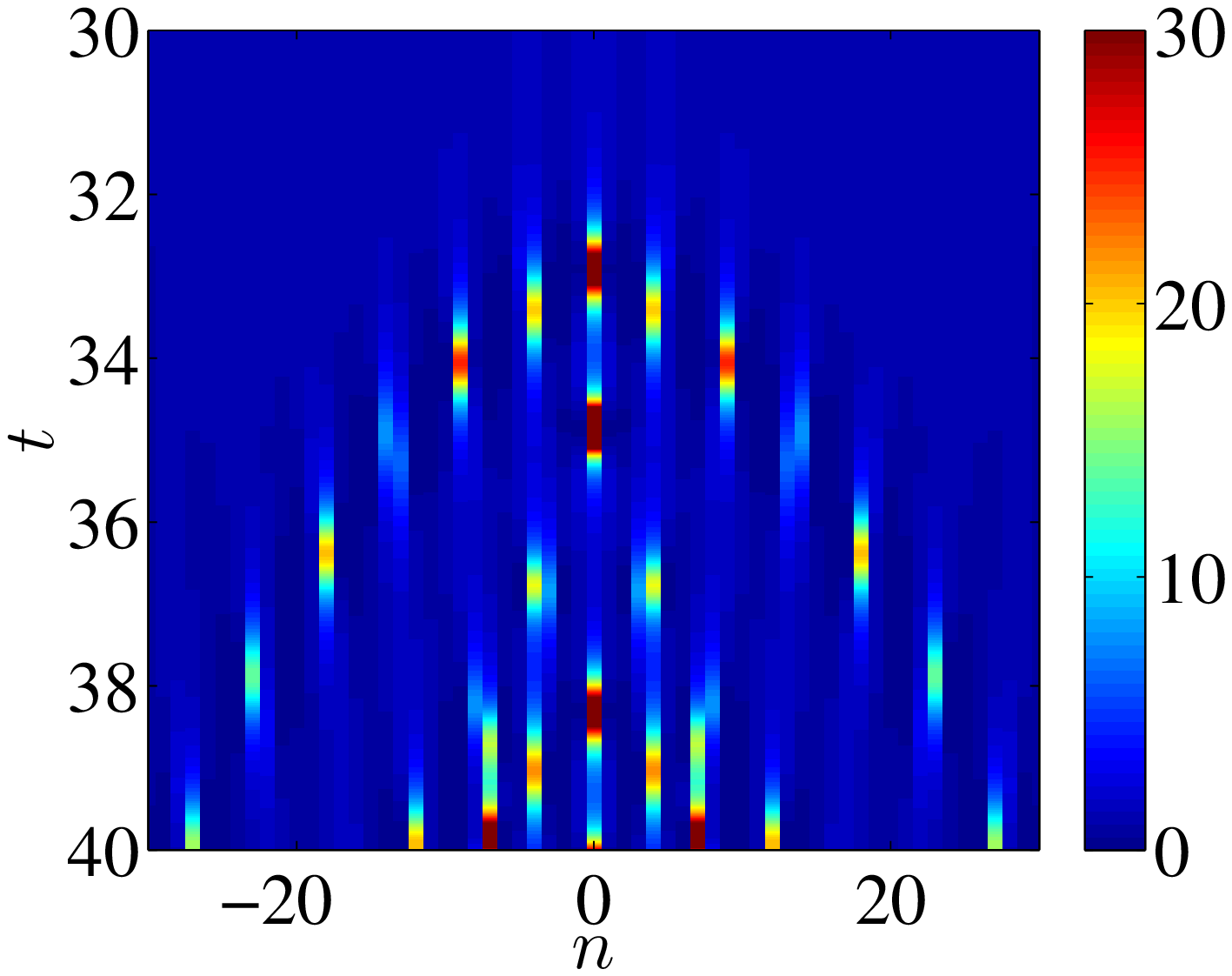}
\includegraphics[height=.18\textheight, angle =0]{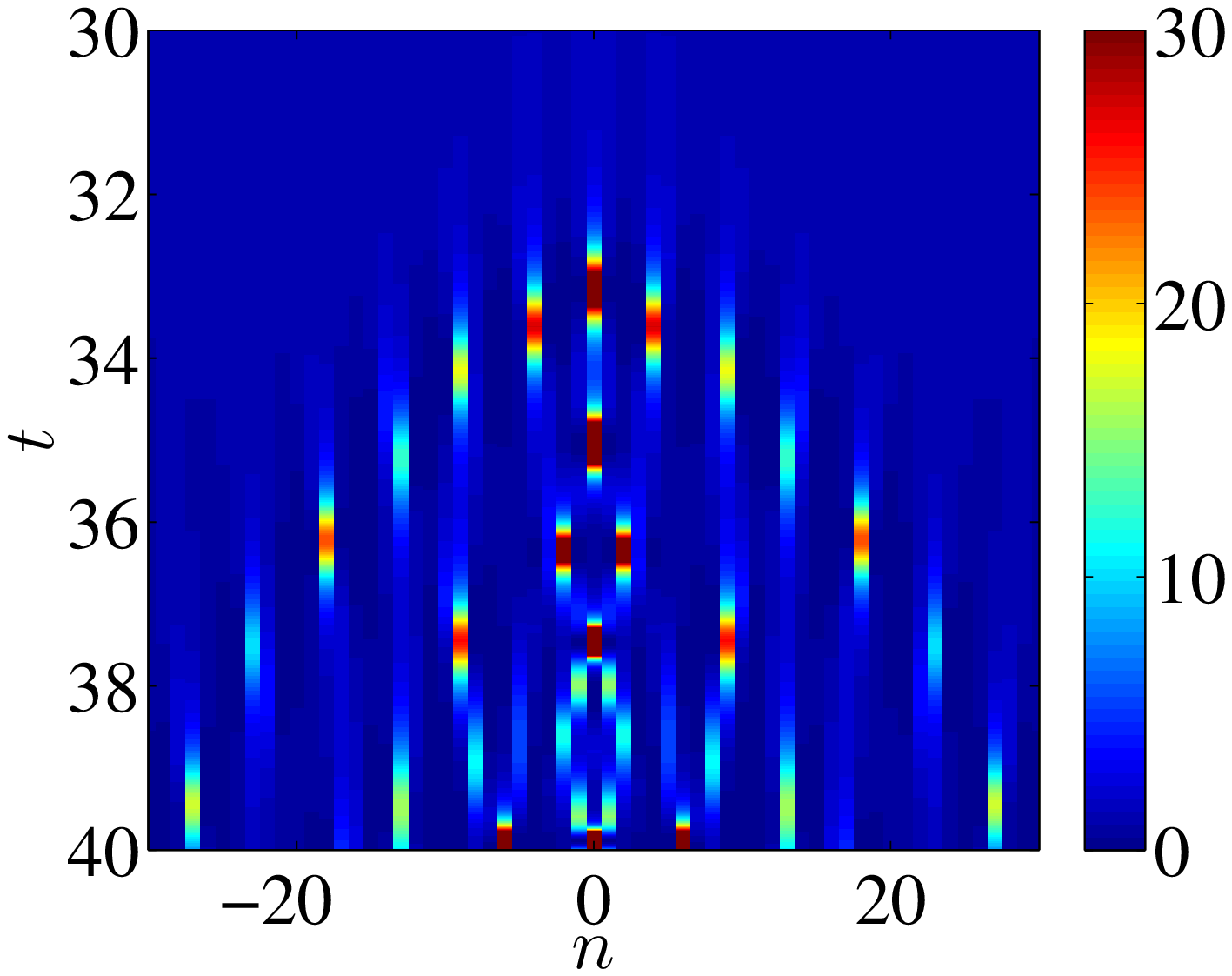}
\includegraphics[height=.18\textheight, angle =0]{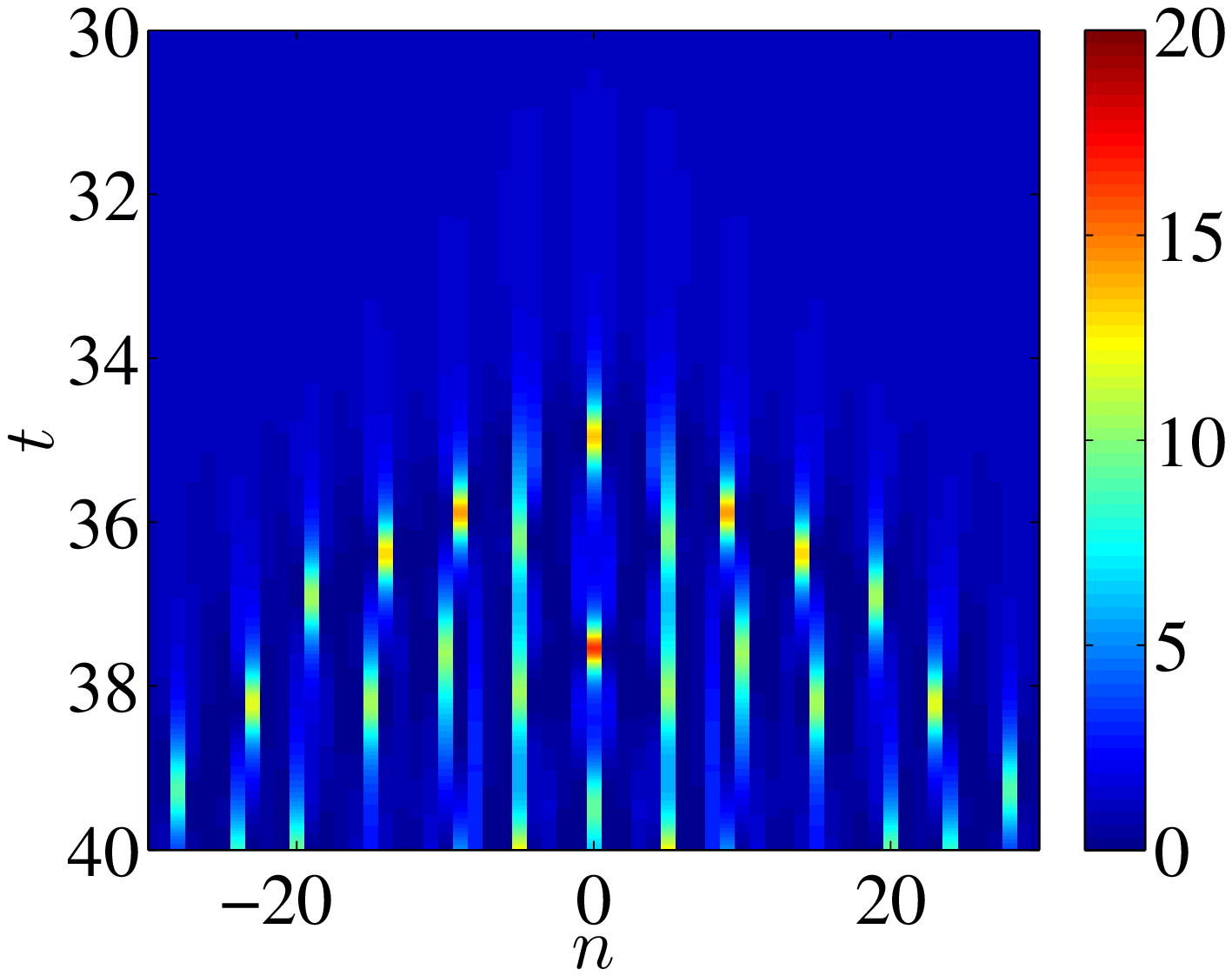}
\end{center}
\caption{
(Color online) Numerical results for the Salerno 
model, 
Eq.~\eqref{gener_model}, with Gaussian 
initial data, for $\mu=0$ (left column, the AL model),
$\mu=0.02$ (center column, weak deviation from AL) and
$\mu=0.5$ (right column, halfway between AL and DNLS).
The first row shows the spatiotemporal evolution of the 
density $|u_{n}|^{2}$ for $\sigma=0.6$, the second for 
$\sigma=1.2$, the third for $\sigma=4.2$ and the last
for $\sigma=18.6$. Similarities (especially in the first
and fourth rows) and differences (especially in the second
and third rows) between the different settings are discussed
in the text.
}
\label{c_panel}
\end{figure}

\end{section}

\begin{section}{Conclusions and Future Work}

  In the present work, we have investigated 
 the formation of   Peregrine soliton-like patterns in 
the far less studied nonlinear dynamical lattices, 
  such as the Ablowitz-Ladik, 
  the discrete nonlinear Schr{\"o}dinger,  
  as well as the Salerno model that interpolates between the two.
  In a similar class of models, and differently 
  than the important statistics-based work of~\cite{tsironis}, we have based
  our investigation on the experience obtained from the continuum
  sibling of the model. This has been done in part due to using
  a (generic) Gaussian initial datum that was shown 
  to give rise in the continuum, in the appropriate semi-classical regime, 
  to Peregrine-type solitons~\cite{stathis}. Even more importantly, it has been
  based on the fundamental mathematical understanding of the
  emergence of the phenomenon of gradient catastrophes and the
  formation during the latter of Peregrine-like structures,
  as per the key work of~\cite{bertola}.

  Indeed, in the discrete integrable model too, relevant Peregrine
  structures (analytically known per the work of~\cite{akhmdis,ohtayang})
  were found to emerge as a result of the gradient catastrophe in the
  semi-classical regime, just like multi-soliton solutions were identified
  away from it. On the other hand though, it was found that even weak
  deviations from the integrable limit would substantially blur the
  picture, and appear to give rise to more permanent/persistent breathing
  patterns which seemed less consonant with the fundamental premise of
  ``appearing out of nowhere and disappearing without a trace''.

  In that light, a key question concerns whether (as we conjectured
  herein) it may be possible to prove that the Peregrine soliton, as a
  homoclinic solution in space-time, no longer
  persists as soon as the discrete, non-integrable perturbation
  comes into play. Performing (perhaps) a Melnikov-type analysis
  and appreciating conditions enabling (or avoiding) the existence
  of Peregrine soliton like waveforms would seem to be an especially intriguing topic for
  future work. Another potentially interesting question might concern
  the relevance of Peregrine patterns for the dynamics of higher-dimensional
  models. For instance, the AL model in higher dimensions is
  known~\cite{almodel} (although it has received very limited
  attention) and moreover it possesses by construction the 1D
  (uniform in the second spatial variable) Peregrine solution.
  A question of interest concerns the ``transverse robustness''
  of this Peregrine and the dynamical outcome of phenomena such as the
  gradient catastrophe in higher-dimensional settings. These
  features are presently under study and will be reported in future
  publications.

\end{section}

\begin{section}*{Acknowledgments}

P.G.K. and D.J.F. acknowledge that this work was
made possible by NPRP Grant No. 8-764-1-160 from Qatar
National Research Fund (a member of Qatar Foundation).
C.H. and P.G.K. gratefully acknowledge that this work was made possible by support from FP7, MarieCurie Actions, People, International Research Staff Exchange Scheme (IRSES-606096). P.G.K. also acknowledges enlightening discussions with
Profs. G. Tsironis and A. Tovbis on the subject.

\end{section}

\end{document}